\documentclass[sigconf]{acmart}
\AtBeginDocument{%
  \providecommand\BibTeX{{%
    \normalfont B\kern-0.5em{\scshape i\kern-0.25em b}\kern-0.8em\TeX}}}

\usepackage{amsmath,amsfonts}
\usepackage{algorithmic}
\usepackage{graphicx}
\usepackage{textcomp}
\usepackage[bottom]{footmisc}
\usepackage{xcolor}
\usepackage{multirow}
\usepackage{balance}       % to better equalize the last page
\usepackage{graphics}      % for EPS, load graphicx instead 
\usepackage[T1]{fontenc}   % for umlauts and other diaeresis
\usepackage{txfonts}
\usepackage{mathptmx}
\usepackage{booktabs}
\usepackage{textcomp}
\PassOptionsToPackage{hyphens}{url}
\usepackage{microtype}        % Improved Tracking and Kerning
\usepackage{ccicons}          % Cite your images correctly!

\usepackage{amsmath,amssymb,amsfonts}
\usepackage{algorithmic}
\usepackage{graphicx}
\usepackage{textcomp}
\usepackage[ampersand]{easylist}
\usepackage{multicol}
\usepackage{multirow}

\usepackage{tabularx}
\usepackage{rotating}
\usepackage[roman]{parnotes}
\usepackage[many]{tcolorbox}
\usepackage{soul}

\usepackage{comment}
\usepackage{booktabs}
\usepackage{verbatim}
\usepackage{array}
\usepackage[inline]{enumitem}
\usepackage[export]{adjustbox}
\usepackage{pifont}

\usepackage{xurl}
\usepackage{booktabs}
\usepackage{hhline}
\usepackage{xcolor}
\usepackage{colortbl}

\setlist[itemize]{noitemsep, topsep=0pt}

\newcommand{\longquote}[2]{
    \begin{itemize}[labelindent=0.65em,labelsep=0cm,leftmargin=*]
    \small
    \item[]{#1: }{\textit{``#2''}} 
    \end{itemize}
  }

\newif\ifdraft
\draftfalse %hides boldifications and comments
\newif\ifdraft
\draftfalse %hides boldifications and comments
\newcommand{\boldification}[1]{\ifdraft\indent ** \textbf{#1} **\\ \indent\else\relax\fi}

\makeatother

\def\BibTeX{{\rm B\kern-.05em{\sc i\kern-.025em b}\kern-.08em
    T\kern-.1667em\lower.7ex\hbox{E}\kern-.125emX}}

\newcolumntype{L}[1]{>{\hsize=.8\hsize\raggedright\arraybackslash}m{#1}}
\newcolumntype{R}[1]{>{\hsize=.8\hsize\raggedleft\arraybackslash}m{#1}}
\newcolumntype{C}[1]{>{\hsize=.8\hsize\centering\arraybackslash}m{#1}}

\copyrightyear{2022}
\acmYear{2022}
\setcopyright{acmcopyright}\acmConference[ICSE-SEIS'22]{Software Engineering in Society}{May 21--29, 2022}{Pittsburgh, PA, USA}
\acmBooktitle{Software Engineering in Society (ICSE-SEIS'22), May 21--29, 2022, Pittsburgh, PA, USA}
\acmPrice{15.00}
\acmDOI{10.1145/3510458.3513009}
\acmISBN{978-1-4503-9227-3/22/05}

\begin{document}

%\def\BibTeX{{\rm B\kern-.05em{\sc i\kern-.025em b}\kern-.08em
%    T\kern-.1667em\lower.7ex\hbox{E}\kern-.125emX}}

%__ end packages and commands from paper____

\title{How to Debug Inclusivity Bugs? A Debugging Process with Information Architecture}

%\author{\IEEEauthorblockN{1\textsuperscript{st} Omitted for Blind %Review}
%\IEEEauthorblockA{\textit{Omitted} \\
%email@email.com}
%}
%\begin{comment}
%
\author{Mariam Guizani}
\affiliation{%
  \institution{Oregon State University}
  \city{Corvallis}
  \state{Oregon}
  \country{USA}}
\email{guizanim@oregonstate.edu}

\author{Igor Steinmacher}
\affiliation{%
 \institution{Northern Arizona University}
 \city{Flagstaff, AZ}
 \country{USA}}
\email{igor.steinmacher@nau.edu}

\author{Jillian Emard}
\affiliation{%
  \institution{Oregon State University}
  \city{Corvallis}
  \state{Oregon}
  \country{USA}}
\email{emardj@oregonstate.edu}

\author{Abrar Fallatah}
\affiliation{%
  \institution{Oregon State University}
  \city{Corvallis}
  \state{Oregon}
  \country{USA}}
\email{fallataa@oregonstate.edu}

\author{Margaret Burnett}
\affiliation{%
  \institution{Oregon State University}
  \city{Corvallis}
  \state{Oregon}
  \country{USA}}
\email{burnett@engr.orst.edu}

\author{Anita Sarma}
\affiliation{%
  \institution{Oregon State University}
  \city{Corvallis}
  \state{Oregon}
  \country{USA}}
\email{anita.sarma@oregonstate.edu}

\renewcommand{\shortauthors}{Guizani et al.}

\begin{abstract}
%\textbf{---------Version 3: ---------}

Although some previous research has found ways to \textit{find} inclusivity bugs (biases in software that introduce inequities), little attention has been paid to how to go about \textit{fixing} such bugs. 
Without a process to move from finding to fixing, acting upon such findings is an ad-hoc activity, at the mercy of the skills of each individual developer.  
To address this gap, we created \textit{Why/Where/Fix}, a systematic inclusivity debugging process whose inclusivity fault localization harnesses Information Architecture(IA)---the way user-facing information is organized, structured and labeled. 
We then conducted a multi-stage qualitative empirical evaluation of the effectiveness of Why/Where/Fix, using an Open Source Software (OSS) project's infrastructure as our setting. 
In our study, the OSS project team used the Why/Where/Fix process to find inclusivity bugs, localize the IA faults behind them, and then fix the IA to remove the inclusivity bugs they had found. 
Our results  showed that using Why/Where/Fix reduced the number of inclusivity bugs that OSS newcomer participants experienced by 90\%.
\\
\\
\textbf{Lay Abstract:}
Diverse teams have been shown to be more productive as well as more innovative. One form of diversity, cognitive diversity --- differences in cognitive styles --- helps generate diversity of thoughts. However, cognitive diversity is often not supported in software tools. This means that these tools are not inclusive of individuals with different cognitive styles (e.g., those who like to learn through process vs. those who learn by tinkering), which burdens these individuals with a cognitive ``tax'' each time they use the tool. In this work, we present an approach that enables software developers to: (1) evaluate their tools, especially those that are information-heavy, to find ``inclusivity bugs''-- cases where diverse cognitive styles are unsupported, (2) find where in the tool these bugs lurk, and (3) fix these bugs. Our evaluation in an open source project shows that by following this approach developers were able to reduce inclusivity bugs in their projects by 90\%.
\end{abstract}
\maketitle
\section{Introduction}
\label{sec:intro}
Although in recent times diversity initiatives have become common, sometimes we forget \emph{why} diversity is important to so many organizations.
Besides social justice reasons, what many organizations hope to gain from diverse backgrounds (cultural, ethnic, education, gender, etc.) is diversity of information and of thought \cite{phillips2014diversity} ---i.e., \textit{cognitive diversity}. 
Diversity's accompanying diversity of thought has been shown to have many positive effects on organizations, including better ability to innovate, better reputation as ethical corporate citizens, and a better ``bottom line'' for businesses~\cite{phillips2014diversity, page2019diversity, larkin2020diversity}.
However, efforts to support diversity rarely consider either cognitive diversity or inclusivity of technology environments. 

In this paper, we consider these aspects together: how to \textit{increase} support for \textit{cognitive diversity} within \textit{technology environments}, especially information-heavy ones.
The setting for our investigation is an information-heavy environment that is particularly challenged in attracting diverse populations: Open Source Software (OSS) communities.

This study complements the existing literature: previous work has investigated OSS-specific challenges ~\cite{steinmacher2015social,guizani2021long,Jensen.King.ea_2011, lee2017understanding} and the inclusivity issues affecting OSS~\cite{Catolino.ea.2019, Robles-2014, Izquierdo:2019, Qiu.ea.2019, mendez2018open, nafus2012patches, bosu2019diversity,padala2020gender, Vasilescu2015}, but has not focused on how to \textit{debug} OSS projects' \textit{technology} to support cognitive diversity. 

\subsection{Why/Where/Fix: An IA-based Inclusivity- Debugging Process}
A debugging perspective suggests that  OSS practitioners who want to improve inclusivity of their project's infrastructure will need three capabilities. 
(1)~First, they need to find ``inclusivity failures'' (analogous to testing \cite{ammann2016introduction}).
Since the failure is about inclusivity (not about producing a wrong output),  OSS practitioners will also need to be able to discern \textit{why} the observed phenomenon is considered an inclusivity failure.
(2)~Second, the practitioners will need to tie an inclusivity failure to \textit{where} the ``inclusivity fault(s)'' occur (analogous to fault localization \cite{avizienis2004basic}); 
so that (3) the inclusivity faults can be \textit{fixed} to stop the associated inclusivity failure from occurring.
In this paper, we term the inclusivity debugging capabilities as  ``Why/Where/Fix'' (Fig. 1), and investigate 
its efficacy at debugging inclusivity bugs.  

\begin{figure}[!b]
\graphicspath{{figures/}}
\vspace{-0.5mm}
\centering
      \includegraphics[width=.47\textwidth]{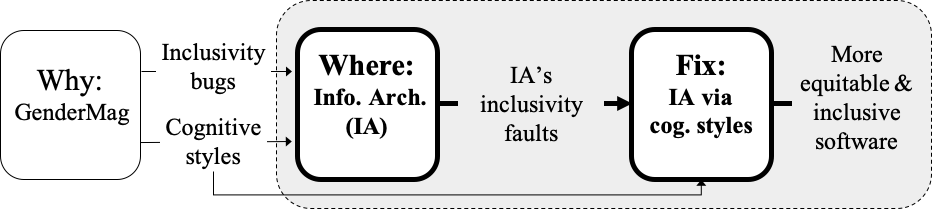}

\caption{The Why-Where-Fix process. The \textit{Why} is to produce the inclusivity bugs and the cognitive styles behind them; the \textit{Where} is to localize the faults behind the bugs to the user-facing IA elements; and the \textit{Fix} is to change the IA elements to expand the software's inclusivity. The grey roundtangle highlights Why/Where/Fix's new contributions.}
\vspace{-4mm}
\label{fig:whyWhereFix}
 \end{figure}

Debugging requires a definition of a bug. We derive our definition from the testing community's notion of a software failure.
Ammann and Offutt define a ``failure'' as
``...external, incorrect behavior with respect to the requirements or ...
expected behavior'' \cite{ammann2016introduction}. 
Our analogous requirement/expected behavior is inclusivity across diverse cognitive styles, so we define \textit{inclusivity failures/bugs} as user-visible features or workflows 
that do not equitably support users with diverse cognitive styles. 
As with Ammann/Offutt's definition, an inclusivity bug is a barrier but not necessarily a ``show-stopper''.  
That is, if groups of users eventually complete their tasks but disproportionately experience barriers along the way (e.g., confusion, missteps, workarounds), these too are inclusivity bugs.

To find such inclusivity bugs and their ``Why''s, we used GenderMag~\cite{burnett2016gendermag}, an empirically validated  method~\cite{Hill2017CHI, padala2020TSE, vorvoreanu-chi19, burnett2017gender} with a dual gender/cognitive focus.
GenderMag integrates finding an inclusivity bug with its ``Why'', because using GenderMag includes identifying \emph{cognitive mismatches} that pinpoint which users disproportionately run into barriers using a system.
In this paper, an OSS team used GenderMag to find inclusivity bugs in their OSS project.

After finding a bug, the next step in debugging is to figure out what and where a bug's causes are, referred to as ``faults'' in SE literature.
According to Avizienis et al. \cite{avizienis2004basic} a fault is the underlying cause of an error, a condition that may lead to a failure; and fault localization is the act of identifying the locations of faults. 
Building upon these definitions, we define an \textit{inclusivity fault} as the user-facing components (e.g., UI elements, user-facing documentation, workflow) of the system that produced an inclusivity bug; 
and \textit{inclusivity fault localization} as the process of identifying the locations of these faults in these user facing components.

Thus, for Why/Where/Fix's ``Where'', we devised a systematic inclusivity fault localization approach that harnesses Information Architecture (IA) \cite{morville2006information}.
IA is the ``blueprint'' for the structure, arrangement, labeling, and search affordances of information content, and is especially pertinent to information-rich environments~\cite{rosenfeld2015information}.
Although substantial research exists on how Information Architectures can support usability, navigation, and understandability~\cite{Rocha.Freixo:2015_IA, Gullikson_IA, Miller_IA_2004, gill2016_IA, lacerda2017information}, research has not considered how different Information Architectures do or do not support populations with diverse cognitive styles, or how IA can be used for inclusivity fault localization.

To use IA to tie together the above ``Why'' and ``Where'' foundations to point to the fixes, we  supplemented the GenderMag process for finding inclusivity bugs with a mechanism by which evaluators specified any IA elements (the faults) implicated in the inclusivity bugs found along the way. 
Thus, the Why/Where/Fix process in Figure~\ref{fig:whyWhereFix}, is: find the bugs using cognitive styles, which contribute the Why (using GenderMag), enumerate the implicated IA elements involved in the bug (Where), and change those IA elements (Fix).

\subsection{Can IA Squash the Inclusivity Bugs?}

We have pointed out that the Why capability (\emph{finding} inclusivity bugs) is already possible using GenderMag.
But debugging requires getting from finding to fixing, and this capability of Why/Where/Fix rests on IA.

Thus, to empirically investigate IA's effectiveness in the Why/ Where/Fix debugging process, we used a three-stage combination of field work (Stage One and Stage Two) and lab work (Stage Three), as follows: 

In Stage~One (Why $\xrightarrow[]{}$ Where), we worked in the field with an OSS team who used GenderMag to detect cognitive inclusivity bugs in their project's infrastructure, to investigate \textit{RQ1: Is IA implicated in inclusivity bugs? If so, how?} 
In Stage~Two (Where $\xrightarrow[]{}$ Fix), the OSS team worked alone to change the project infrastructure's IA using what they had learned in Stage One, which enabled us to investigate \textit{RQ2: Can practitioners use IA to fix inclusivity bugs? If so, how?}  
In Stage~Three (Lab), we brought OSS newcomers into the lab to investigate whether the team's IA-localized faults and fixes decreased the inclusivity bugs those newcomers experienced.

Our primary contributions in this paper are: 
\\(1) Presents and empirically investigates the first inclusivity debugging process, including systematic fault localization.
\\(2) Empirically investigates whether Information Architecture can itself be the cause of inclusivity bugs.
\\(3) Reveals ways OSS projects can improve their infrastructures' Information Architecture to improve their project's inclusivity.

\section{Background and Related Work }

\subsection{Information Architecture}
\boldification{IA is defined as ABC. It describes how the information is organized to provide the users a better experience. We follow Rosenfeld defn.}

\begin{figure}[!b]
\graphicspath{{figures/}}
\centering
      \includegraphics[width=.47\textwidth]{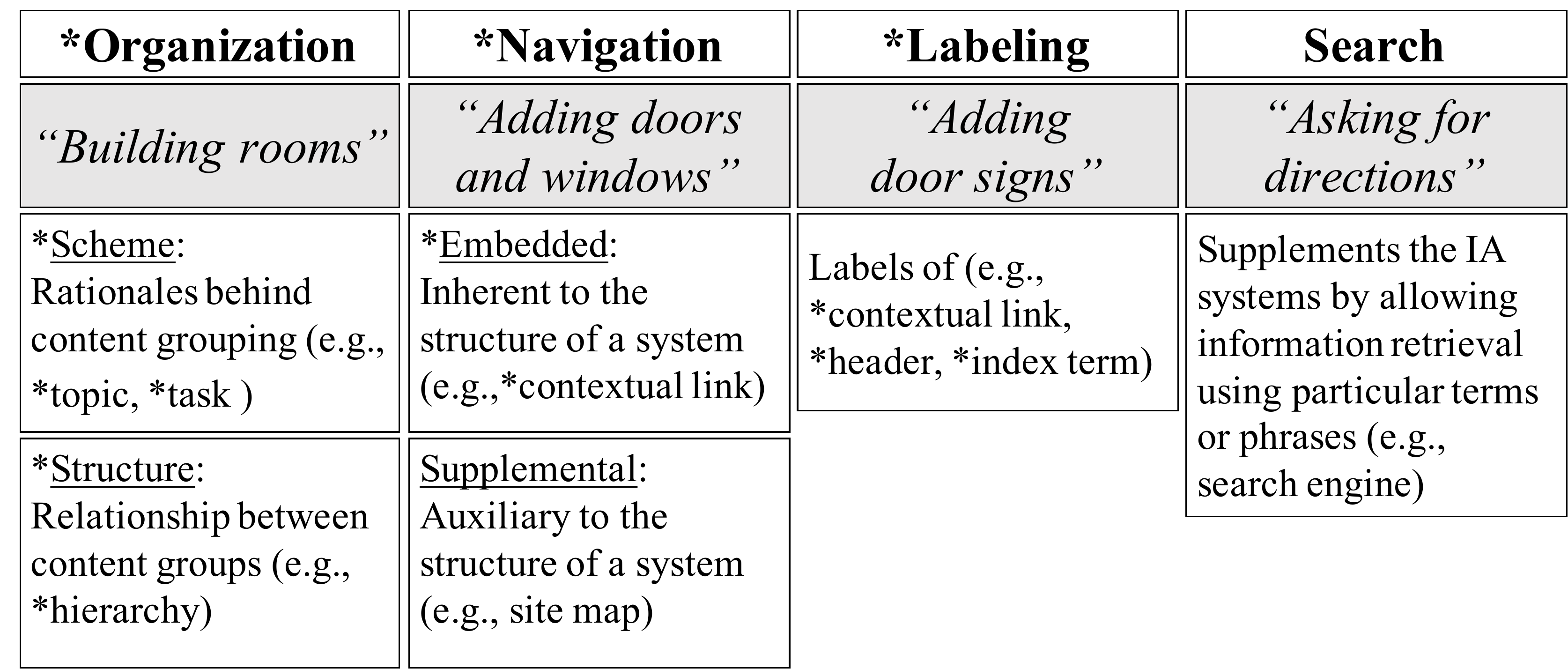}
\caption{IA's four component systems~\cite{morville2006information}. The organization and navigation systems have subsystems (underlined). *s mark IA (sub)systems and elements used in this paper.
}
\vspace{-2mm}
\label{fig:MorvilleIAsystems}
 \end{figure}

The term ``Information Architecture'' was coined in the mid-70's as a way of ``making the complex clear'' \cite{wurman1975infoArchitecture}.
This paper follows the definition of Morville and Rosenfeld~\cite{morville2006information}, referred to as the ``bible'' of IA, that defines IA as a set of four component systems (Figure~\ref{fig:MorvilleIAsystems}).

The first is the \textit{Organization System (Org)}, analogous to the architectural arrangement of a building's ``rooms'', which has an organization scheme \textit{OrgScheme} and an organization structure \textit{OrgStruct}.
The organization scheme is the way content is arranged or grouped (e.g., alphabetical or by task). 
An architect chooses the scheme according to the situations they want the Information Architecture to support, such as alphabetical \textit{(OrgScheme-Alpha)} to support exact look-ups, or task-based \textit{(OrgScheme-Task)} to facilitate high priority tasks. 
The organization structure defines the relationship between content groups (e.g., hierarchical \textit{(OrgStruct-Hierarchy)}).

Second, the \textit{Navigation System (Nav)}, analogous to adding doors and windows to a building, enables users to traverse the information groupings and structure. 
Some of the navigation system is embedded in the content (e.g., contextual links \textit{(Nav-ContextualLink)}), while others are supplemental (e.g., site maps).
Third, the \textit{Labeling System} (Label) adds signposts (also known as ``cues'' in Information Foraging literature \cite{pirolli2007information}) to the ``doors'', such as the labels on contextual links \textit{(Label-ContextualLink)}, headers  \textit{(Label-Header)}, cues/keywords \textit{(Label-IndexTerm)}, etc. 
Fourth, the \textit{Search System}, when provided, supplements the rest of the IA, to enable users to retrieve information using a particular term or phrase. 

\boldification{IA is widely used.}
The majority of IA research has focused on the design and evaluation of websites, but some has explored other domains.
For example, IA has been used in the design of usable security tools \cite{de2019smart}, as the basis of a semantic web structuring tool~\cite{garcia2011publishing, brunetti2012improved, brunetti2013design}, to investigate the accessibility, use and reuse of information across multiple devices~\cite{oleksik2013towards}, to evaluate different information visualization tools~\cite{li2017comparing}, and for mobile applications screen-reader navigation~\cite{gross2018exploring, yang2012aural}. 
\boldification{Some research looked at IA in general comparing it with other usability concepts ...} 
One body of research has compared IA to other attributes of information sites. 
For example, Aranyi et al.'s empirical evaluation of a news website showed that the content and its IA were the main problems~\cite{aranyi2012using}. 
Petri and Power's study likewise found prominent IA problems when evaluating six government websites, with IA accounting for about 9\% of user-reported bugs~\cite{petrie2012users}. 

\boldification{Others looked focused on some subsystems of Information Architecture  ...} 
Other IA research has evaluated the usability of different subsets (organizational vs. labeling schemes) of IA.
For example, Gullikson et al. evaluated the IA of an academic website and reported that although participants were satisfied with the content of the site, they found its (IA) labeling to be confusing  \cite{gullikson1999impact}, and were especially dissatisfied with the IA's organization system.
Resnick and Sanchez found that user-centric labels significantly improved user performance and satisfaction as compared to user-centric organization, which only improved performance if labels were of low quality~\cite{resnick2004effects}. 
Similarly, others have found that navigation success depends more on the quality of labels than the structure of a page \cite{miller2004modeling, schaik2015automated}.

\boldification{Of particular interest to this paper is the IA research pertaining to the support of diverse populations, and here's what we have about that}
Of particular interest is IA research on supporting diverse populations.
Lachner et al. used IA to promote cultural diversity and used Hofstede et al. power distance cultural dimension   \cite{hofstede2011dimensionalizing} to design and evaluate culturally-specific collaborative Q\&A websites \cite{lachner2018culturally}.
Accessibility and IA has been studied by others.
Swierenga et al. showed that IA's organization and labeling system create barriers for visually impaired and low-vision individuals~\cite{swierenga2011website}.
A multitude of research \cite{bolchini2006designing, gatsou2012novice, rohani2013mobile, rohani2012back, vigo2013challenging, yang2012aural} has investigated IA auditory systems for designing and evaluating accessible websites for low-vision users. Ghahari et al., for example, showed how topic- and list-based aural navigation strategies can enhance user's navigation effectiveness and efficiency~\cite{rohani2012back}.
However, we cannot locate any research on how IA can support cognitive diversity.
%__________________________________________

\subsection{Diversity and the GenderMag Method}
\label{subSec:GenderMag}

\boldification{GenderMag is... and has facets embedded in 3 multi-personas.}
GenderMag, a method used to find and fix inclusivity bugs, provides a dual lens---gender- and cognitive-diversity---to evaluate workflows.
It considers five dimensions (``facets'' in GenderMag) of cognitive styles (Table \ref{table:GenderMagCodeset}),
each backed by extensive foundational research~\cite{burnett2016gendermag, stumpf2020gender}.
Each facet has a range of possible values.
A few values within each facet's range are brought to life by the three GenderMag personas: ``Abi'', ``Pat'', and ``Tim.'' 
Abi's facets are statistically more common among women than other people, Tim's are statistically more common among men, and Pat has a mix of Abi's/Tim's facets plus a few unique ones. 

\boldification{The personas are ''multi-'' in that... see Figure Abi}

\begin{figure}[!b] 
\graphicspath{ {figures/} }
\vspace{2mm}
\centering
         \includegraphics[width=.47\textwidth]{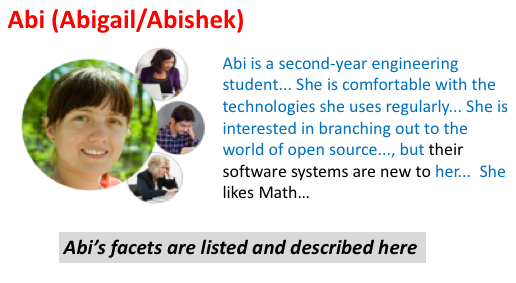}
\vspace{-4mm}
\caption{Portions of the OSS team's Abi persona. The photo(s) and \textcolor{cyan}{blue} text are customizable; the black text is not. Abi's facets (gray block) are as per Table~\ref{table:GenderMagCodeset}. (The supplemental document \cite{suppdoc} includes the full Abi persona used in Stage~One.)}
\label{fig:Abi_background}
\vspace{-3mm}
\end{figure}

Each persona is a ``multi-''persona~\cite{Hill2017CHI}---their demographics can be customized to match those of the system's target audience.
For example, any gender, any photo, any educational background, or any pronoun can be integrated (e.g., she/her, he/him, they, ze, etc.).
Their cognitive facets, however, remain fixed.
Figure \ref{fig:Abi_background} shows portions of the OSS team's customization of Abi, which they used in Stage~One.)
\begin{table}[!b] %was [t!]
\centering
\caption{The GenderMag cognitive facet values for each persona. The research behind each facet is enumerated in~\cite{burnett2016gendermag}. %\textcolor{red}{**it would be awesome if this table showed up WITH the Abi figure}
}
\vspace{-3mm}
\begin{tabular}{L{1cm}L{8.8cm}}
\hline
{\textbf{Facet}}
& { \textbf{Cognitive facet value for each persona}}\\\hline

\rotatebox{90}{ ~Motivations~}
&  Uses technology...
\emph{Abi}: Only as needed for the task at hand. Prefers familiar and comfortable features to keep focused on the primary task.\newline
\emph{Tim}: To learn what the newest features can help accomplish.\newline
\emph{Pat}: Like Abi in some situations and like Tim in others.
\\ \hline

\rotatebox[origin=c]{90}{  \parbox{1cm}{\centering Self-Efficacy}}
&   
\emph{Abi}: Lower self-efficacy than their peers about unfamiliar computing tasks. If tech problems arise, often blames self, and might give up as a result. \newline
\emph{Tim}: Higher self-efficacy than their peers with technology. If tech problems arise, usually blames the technology. Sometimes tries numerous approaches before giving up. \newline
\emph{Pat}: Medium self-efficacy with technology. If tech problems arise, keeps on trying for quite awhile. 
\\\hline

\rotatebox[origin=c]{90}{  \parbox{1cm}{\centering Attitude\\Toward Risk}}
&   
\emph{Abi} and \emph{Pat}: Risk-averse, little spare time; like familiar features because these are predictable about the benefits and costs of using them. \newline
\emph{Tim}: Risk tolerant; ok with exploring new features, and sometimes enjoys it.  
\\ \hline

\rotatebox[origin=c]{90}{  \parbox{1.5cm}{\centering Information\\Processing}}
&   
\emph{Abi} and \emph{Pat}: Gather and read everything comprehensively before acting on the information. \newline
\emph{Tim}: Pursues the first relevant option, backtracking if needed.
\\ \hline

\rotatebox[origin=c]{90}{ \parbox{1cm}{\centering Learning\\Style}}
&   
\emph{Abi}: Learns best through process-oriented learning; (e.g., processes/algorithms, not just individual features).\newline
\emph{Tim}: Learns by tinkering (i.e., trying out new features), but sometimes tinkers addictively and gets distracted. \newline
\emph{Pat}: Learns by trying out new features, but does so mindfully, reflecting on each step. 
\\\hline
 
\end{tabular}
\label{table:GenderMagCodeset}
%\vspace{-5mm}
\end{table}

\boldification{The way you do a Gendermag eval is like this...}
Evaluation teams, such as the OSS team in this paper, use GenderMag to walk through a use-case in the project they are evaluating using Abi, Pat, or Tim.
At each step of the walkthrough, the team writes down the answers to three questions: (1)~whether <Persona> would have the subgoal the project owners hoped for and why, (2)~whether <Persona> would take the action the project owners hoped for and why, and (3)~if <Persona> did take the hoped-for action, would they know they did the right thing and were making progress toward their goal, and why.
When the answer to any of these questions is negative, it identifies a potential bug; if the ``why'' relates to a particular cognitive style, this shows a disportionate effect on people who have that cognitive style---i.e., an \textit{inclusivity bug}.
Thus, a team's answers to these questions become their inclusivity bug report, which they can then process and prioritize in the same way they would do with any other type of bug report.

\boldification{**GM is pretty accurate, and it's pointed out a lot of problems in OSS before...}

The method and its derivatives have been used in a variety of domains, such educational software, digital libraries, search engines, and software tools~\cite{gralha2019, burnett2016gendermag, cunningham2016gendermag, hilderbrand2020gendermag-bp, shekhar2018gendermag,mendez2018open,vorvoreanu-chi19}.
Particularly pertinent to this paper, in a study of OSS professionals, over 80\% of the barriers they found in OSS projects were gender inclusivity bugs, which were later confirmed by OSS newcomers~\cite{padala2020TSE}.

\boldification{but nobody has looked at inclusivity FL before, so that's what we're doing, using IA}
However, prior work has left largely to the practitioners' judgment how exactly to fix such inclusivity bugs (e.g.,~\cite{vorvoreanu-chi19}).
This paper aims to pave a path from finding to fixing with an IA-based process to systematically localize inclusivity faults.

\section{Methodology}
\boldification{Can a more inclusive IA make a difference?To find out, we worked with Project~F...}
We conducted a multi-stage (in-the-field and in-the-lab) empirical investigation to analyze whether changing the IA of an OSS project infrastructure would help support newcomers across a range of diverse cognitive styles.\footnote{We did not \textit{recruit} participants with any particular cognitive style as a criterion; rather, we \textit{collected} cognitive style data as part of the investigation.}  

For the field aspect, we gathered in-the-field data from an OSS project team (Team~F) that was interested in increasing diverse newcomers' participation in their project (Project~F).
The empirical investigation had three stages: 
\begin{itemize}[labelindent=0.3em,labelsep=0.2cm,leftmargin=*]
    \item Stage~One (Why $\xrightarrow[]{}$ Where), in the field: We worked with Team~F to detect IA-based inclusivity bugs. Team~F then worked alone to select which of these bugs to fix.
    \item  Stage~Two (Where $\xrightarrow[]{}$ Fix), in the field: Team~F worked alone to derive IA-based cognitive diversity-inspired fixes to Project~F's Information Architecture. 
    \item Stage~Three, in the lab: We brought OSS newcomers into the lab to evaluate the inclusivity bugs they encountered with the original Project~F vs. the new version of Project~F.
\end{itemize}

%---------------------------
\subsection{Stage 1, Team F, RQ1 (in the field): \\Why $\xrightarrow[]{}$ Where}

Stage~One had two purposes.
First, for ecological validity, we wanted to avoid artificially creating inclusivity bugs; thus, Stage~One provided a way to harvest them from a real OSS project.
For this purpose, we used the GenderMag method (Section \ref{subSec:GenderMag}). 
To facilitate IA-based fault localization, we then added the following IA-based Where question to the GenderMag question set: ``What in the UI helped/confused <Persona> in this step?''
Both the original and IA-supplemented GenderMag forms, and all our study materials, are provided in the supplemental document~\cite{suppdoc}.

Note that Stage~One's purpose was \emph{not} to investigate whether an OSS team can use GenderMag to  
point out inclusivity bugs, because its validity with OSS project teams has already been validated~\cite{padala2020TSE}.
The GenderMag method has also been empirically validated in other lab~\cite{burnett2016gendermag, Hill2017CHI, vorvoreanu-chi19} and  field~\cite{burnett2017gender} studies. 
As with other cognitive walkthrough (CW) methods, its reliability (precision) is very high: CW methods tend to have false-positive rates of 5\%-10\%, and GenderMag's false-positive rates have been 5\% or lower~\cite{burnett2016gendermag,  vorvoreanu-chi19, padala2020TSE}.

For Stage~One's bug harnessing purpose, Team~F worked with two researchers using the IA-supplemented GenderMag method to find inclusivity bugs in four use-cases (Table~\ref{table:usecasesdescriptions}). 
Team~F selected these use-cases for their importance for Project~F newcomers.
Analyzing these use-cases produced both a list of likely inclusivity bugs with the facets that caused them (Why), and IA-localized faults that may have produced these bugs (Where).

The second purpose of Stage~One was the beginning of our RQ1 investigation into whether some IA elements are indeed implicated in such real-world inclusivity bugs.
For this purpose, Team~F worked alone, without our help. 

Team~F began by deciding which of the bugs to take forward into the next stage of the investigation.
They selected these bugs using the criteria that the bug (1)~had at least one cognitive facet that the Information Architecture did not support; and (2)~was associated with the project itself and \textit{not} the UI of the hosting platform (e.g., GitLab, GitHub). 
These criteria produced 6 bugs (Table \ref{table:usecasesdescriptions}).

Along the way, Team~F had noticed some general usability bugs not related to any cognitive facet. To prevent these from influencing Stage~Three, Team~F fixed these bugs and brought the project up to GitHub's recommended content standards~\cite{OSSGuidelines}, resulting in the prototype we call the \emph{Original} version. 

\begin{table}[!b]
\vspace{-0.2cm}
\caption{The four use-cases and associated bugs. Team~F provided these use-cases, which were  important to their project.}
\vspace{-2mm}
\label{table:usecasesdescriptions}

% \
\begin{tabular}{L{2.4cm}L{5.1cm}L{1.7cm}}
\hline
    { \textbf{Use-Case}}
    & {\textbf{Descriptions}}
    & {\textbf{Bugs}}
    \\\hline

%\hangindent=1em
%\hangafter=1
     \ U1-Find
     & \ Finding an issue to work on
     & \ Bug 1 \& 2 
\\%\hline

%\hangindent=1em
%\hangafter=1
     \ U2-Document
& \ Contribute to the documentation & \ Bug 3 
    \\%\hline

%\hangindent=1em
%\hangafter=1
     \ U3-FileIssue
     & \ File an issue
     & \ Bug 4
    \\%\hline

     \ U4-Setup
     & \ Set up the environment
     & \ Bug 5 \& 6
    \\\hline

%\hangindent=1em
%\hangafter=1

\end{tabular}

\vspace{-4mm}
\end{table}

%---------------------------
\subsection{Stage 2, Team F, RQ2 (in the field): \\Where $\xrightarrow[]{}$ Fix }
   
Team~F then worked alone to derive fixes for each of these 6 bugs by changing the IA elements they had identified as the probable causes of the bugs, so as to better support the previously unsupported cognitive facets without loss of support for the supported facets.
We refer to the ``fixed'' version of Project~F as the \emph{DiversityEnhanced} version.

%---------
\subsection{Stage 3, OSS Newcomers, RQ1+RQ2 (in the lab)}
\label{subsec:stageThree}
\boldification{Did these fixes work?}
We then brought OSS newcomers into the lab to investigate: \linebreak (1) whether OSS newcomers trying to use the Original version would run into the bugs Team~F had found in the Original version, and (2)~whether the IA fixes Team~F had derived for the DiversityEnhanced version would 
actually improve support for cognitively diverse OSS newcomers.

\boldification{Who were our user study participants and how did we recruit them?}

We recruited the OSS newcomers from a large US university.
Our recruiting criteria were people with no prior experience contributing to OSS projects. 
All 31 respondents came from a variety of science and engineering majors. 
Because the investigation focuses only on cognitive diversity (not on disabilities), we did not seek out participants with any particular cognitive style or with a disability.
Because none of the experimental tasks required programming, we did not collect their programming experience.

\begin{figure}[!b]
\vspace{0.7em}
\graphicspath{ {figures/} }
\centering
         \includegraphics[width=.33\textwidth]{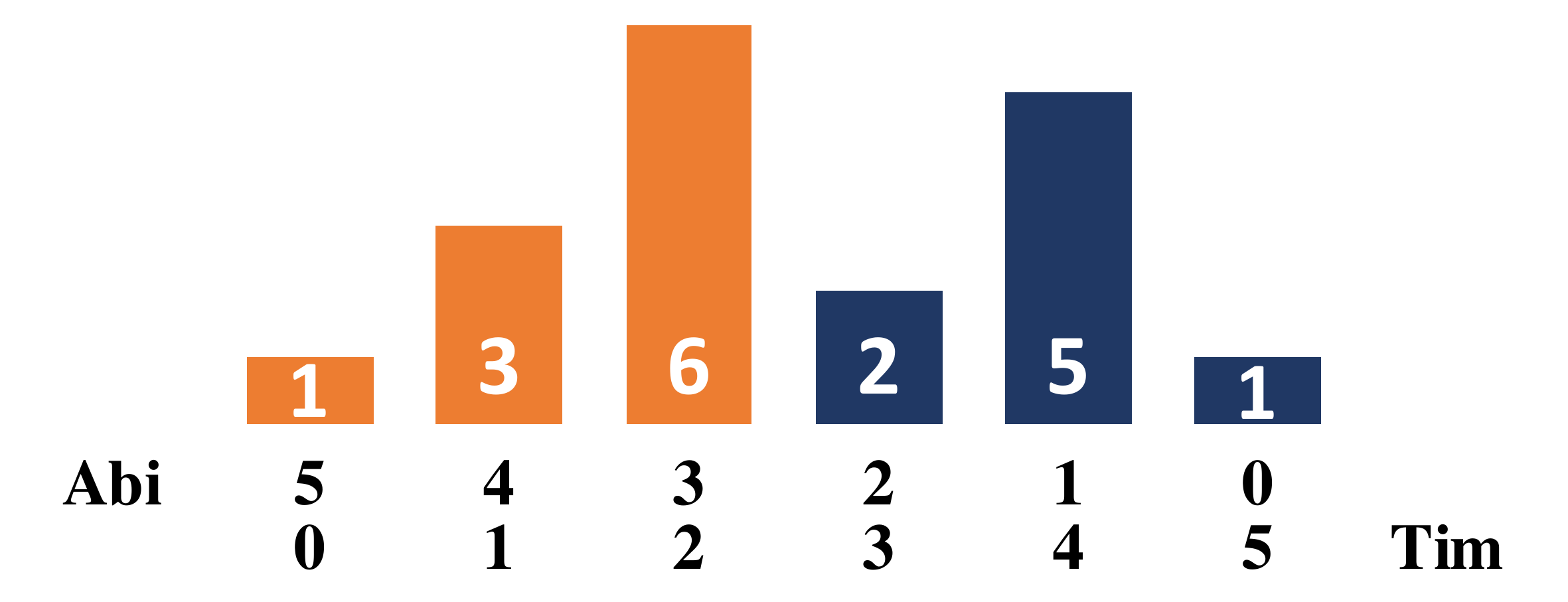}
\caption{Number of participants with more Abi facets (left half, orange) or more Tim (right half, blue). For example: the first column says that 1 participant had 5 Abi facets and no Tim facets. Table~\ref{table:GenderMagCodeset} explains Abi, Tim, and their facets. } 
\label{fig:PersonaDistribution}
\end{figure}

Participants filled out a cognitive facet questionnaire~\cite{burnett2017gender, vorvoreanu-chi19, gralha2019} (provided in our supplemental document \cite{suppdoc}) in which participants answered Likert-scale items about their cognitive styles.
Using their responses and genders, we selected 18 respondents to gender-balance and to include a wide range of cognitive styles (Figure~\ref{fig:PersonaDistribution}). 
Of the 18 selected participants, 8 identified as women, 9 identified as men, and one participant declined to specify their gender. 

\begin{figure}[!b]
\vspace{-2mm}
\graphicspath{ {figures/} }
\centering
         \includegraphics[width=.47\textwidth]{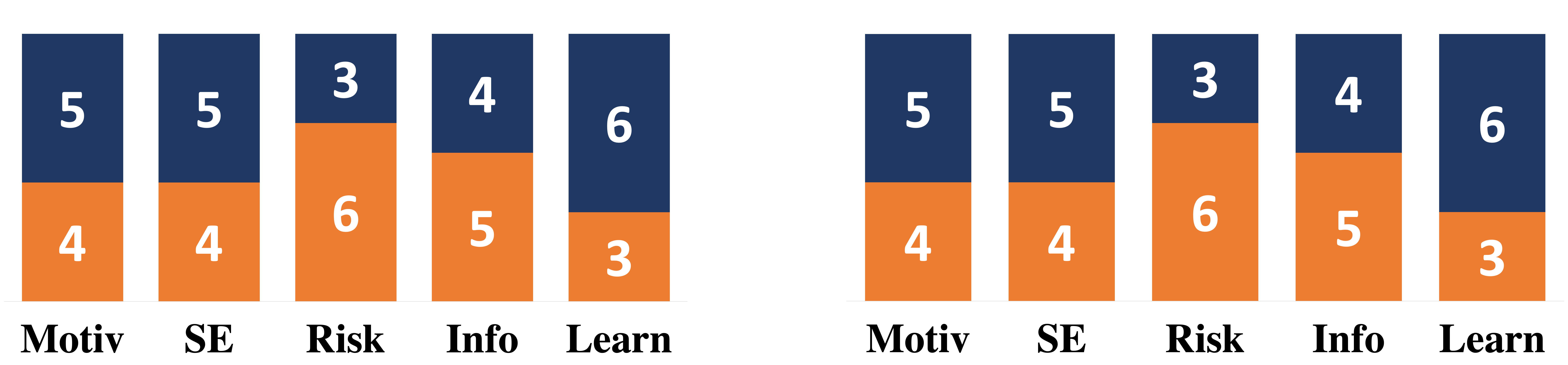}
\caption{Number of participants with Abi (bottom, orange) vs. Tim (top, blue) facets who used the Original (columns 1-5) vs. DiversityEnhanced (columns 6-10) versions of the OSS project. (The two distributions are identical.) }
\vspace{-2mm}
\label{fig:facetpergroup}
\end{figure}

\boldification{who went to which group?}
We assigned participants to the Original or DiversityEnhanced treatments, balancing the cognitive styles between the treatments based on the participants' cognitive facet questionnaire responses.
Because facet values are relative to one's peer group, the median response for each facet served to divide closer-to-Abi facet values from closer-to-Tim facet values.
This produced identical facet distributions (Figure~\ref{fig:facetpergroup}) for both groups. 

We audio-recorded each participant as they talked-aloud while working on the use-cases presented earlier in Table~\ref{table:usecasesdescriptions}. 
We transcribed the recordings, and counted how often the participants encountered one of the 6 bugs that Team~F had attempted to fix.

We qualitatively coded cognitive facets that participants verbalized when they encountered one of these bugs, which enabled us to compare participants'
\textit{in-situ} reactions to their cognitive facet questionnaire responses. For example, we coded P2-O's verbalization \textit{``...this leads me to a page with the bare minimum of instructions... I have no idea where to go from here''} as ``learning style: process-oriented'', which aligned with their questionnaire response.
To ensure reliability of the coding, two researchers independently coded 20\% of the data and calculated IRR using the Jaccard index. Jaccard, a measure of ``consensus'' interrater reliability \cite{stemler2004}, is useful when multiple codes per segment are used, as in our case. 
The consensus level was 90.2\%. 
Given this level of consensus, the researchers split up coding the remainder of the data. 

At session end, participants filled out a subset of the System Usability Scale (SUS) survey~\cite{brooke1996sus} (supplemental document \cite{suppdoc}). 
\section{Results}
\label{sec:results_study2and3}

\boldification{The goal of this section is to answer RQ1 and RQ2}
We begin with ``whether'' answers to both research questions---for RQ1, \textit{whether} Information Architecture was implicated in the inclusivity bugs, and for RQ2, \textit{whether} Team~F's IA fixes increased inclusivity for OSS newcomers.

\boldification{and the short answer is yes... and this is how... }
As Table~\ref{table:UserStudyGeneral} shows, both answers were yes.
Regarding RQ1, with the Original version, OSS newcomers ran into inclusivity bugs in the Information Architecture 20 times. 
Regarding RQ2, Team~F's inclusivity fixes to the IA reduced the number of inclusivity bug experiences in the DiversityEnhanced version to only 2. In total, Team~F's IA fixes cut the number of bugs participants experienced by 90\% (Table~\ref{table:UserStudyGeneral}). 

\boldification{and this is how... }
To answer the \textit{how} aspects of our RQs, Table~\ref{table:FlossIssuesFixes} \textcolor{black}{summarizes, for each bug,  Team~F's \textit{Why} analyses (first column) of the cognitive facets involved in the bug, their \textit{Where} analyses to localize the faults to IA elements (second column), and how they implemented their IA \textit{Fixes} (third column). 
The following sections discuss them in depth. 
} 

\begin{table}[!b]
  \caption{The number of participants who ran into the bug(s), out of the 18 participants (9/ group). %Issues in the same use-case are grouped together except for Issues 4 \& 6, which each required separate scoring for information content (Section~\ref{subsec:stageThree}).
  } %The last row totals the number of issues reported by the participants.}% participants who ran into each issue(s).
  \vspace{-2mm}

\begin{tabular}{L{4.25cm}C{2cm}C{3.25cm}}
  \hline
   {   \textbf{Bug ID} }
    & {  \textbf{Original }}
    & {  \textbf{DiversityEnhanced} }
    \\\hline

    Bug 1 \& 2 &   9/9 &   1/9 \\ %\hline  

    Bug 3&   2/9 &   0/9 \\ %\hline  
  
  %  Issue 4&   9/9 &   9/9 \\ %\hline  
  
    Bug 4&   0/9 &   0/9 \\ %\hline
  
  %  Issue 6&   9/9 &   7/9 \\ %\hline  

    Bug 5 \& 6 &   9/9 &   1/9 \\ %\hline  
   \textbf{Total bugs encountered} &   \textbf{20} &   \textbf{2}\\ 
 %  \textbf{Mean}}&   \textbf{5}} &   \textbf{1.75}}\\ 
 \hline
  \end{tabular}

  \label{table:UserStudyGeneral}

\end{table}
\begin{table*}[!ht]
    \caption{For each use-case's bug(s), excerpts from Team~F's Stage~One analysis, the Bug's Why's (facets impacted), Where's (IA involved), and their Stage~Two IA fixes.}  
    \vspace{-2mm}
    \centering
    \includegraphics[width=\textwidth]{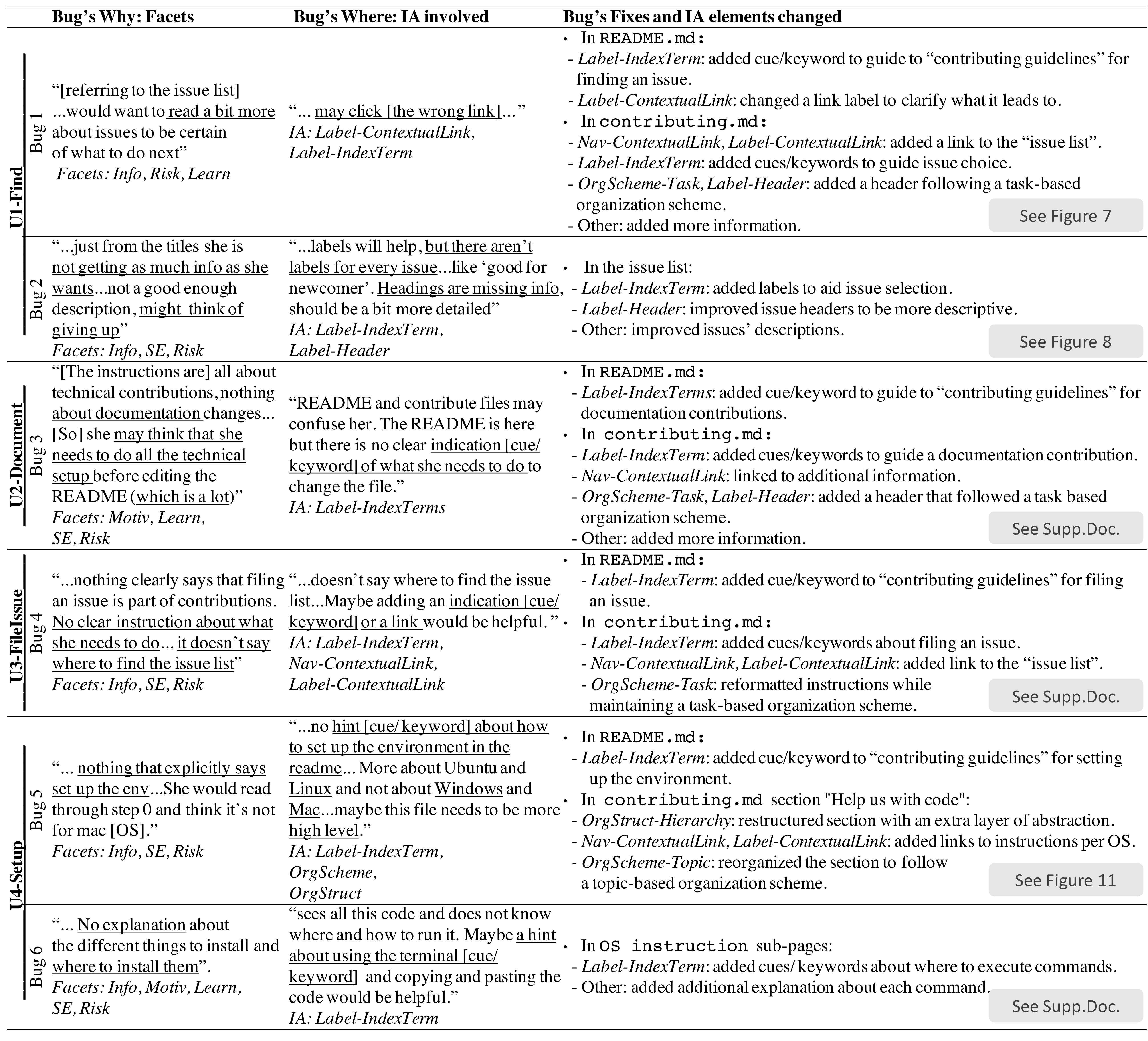}
\label{table:FlossIssuesFixes}
\end{table*}
%_________________________________________

\subsection{Bug 1 \& 2 in Depth: Issues with the ``issue list'' }

\textcolor{black}{The first two rows in Table~\ref{table:FlossIssuesFixes} show how Team~F addressed Bug~1 \& 2, the IA-based inclusivity bugs that Team~F identified in Stage~One in the context of use-case U1-Find (finding a task to work on).
}

\boldification{Bug 1: what, why and where...}
\textcolor{black}{As Table~\ref{table:FlossIssuesFixes} shows, }
for Bug~1, Team~F predicted that Abi-like newcomers would face problems in understanding the \textit{process} of finding an issue.

Their \textit{why} analysis (Table~\ref{table:FlossIssuesFixes} row 1 col. 1) pointed out that the lack of information about finding an issue could be problematic to comprehensive information processors, risk averse, or process-oriented newcomers.
As \textcolor{black}{Stage~Three} \underline{P}articipant \underline{1} using the \underline{O}riginal version later put it: 
\longquote{P1-O}{I just feel like I wouldn't have enough to go on.}

\textcolor{black}{Team~F localized the fault (\textit{wheres}, Table~\ref{table:FlossIssuesFixes}'s row 1 col. 2) to the IA's link labeling \textit{(Label-ContextualLink)}
and to 
} 
the absence of keywords \textit{(Label-IndexTerm)}, 
\textcolor{black}{which could lead newcomers to follow wrong link(s) and never obtain the kind of information they were seeking.}  

\boldification{and here's the what, why and where for Bug~2}
\textcolor{black}{Once a newcomer was past Bug~1, Team~F predicted that the Issue List provided too little information to enable some newcomers to select an issue appropriate to their skills (Bug~2).
Team~F's \textit{why} analysis showed that this bug would be particularly pertinent to
newcomers with a comprehensive information processing style, low self-efficacy, or risk aversion. 
} 

\begin{figure}[h!t]
\graphicspath{ {figures/} }
\centering
         \includegraphics[width=.47\textwidth]{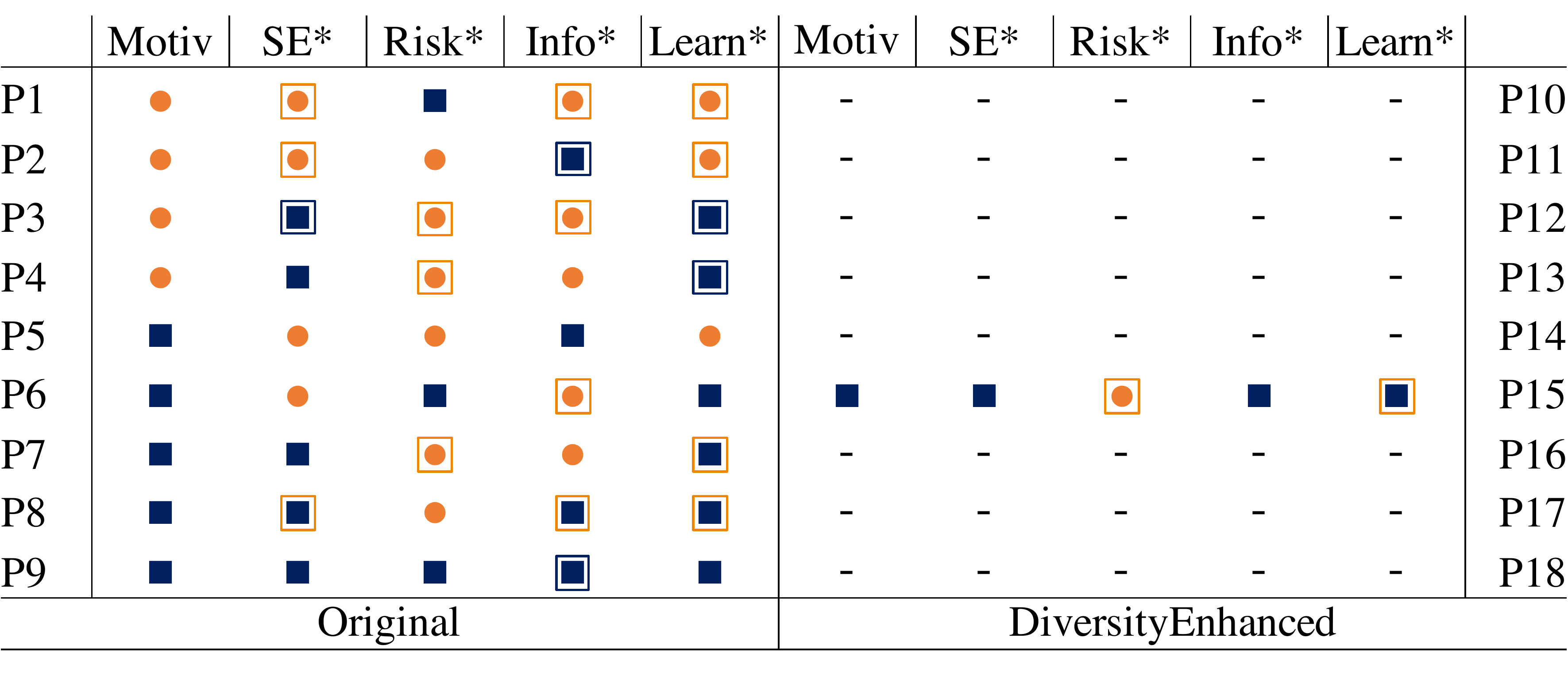}
\caption[caption]{In Bug 1 \& 2, all Original participants ran into bugs (left), but only 1 DiversityEnhanced participant (right). Participant ID numbering is from the most Abi-like to the most Tim-like.
\\ *: facet the fix(es) targeted;
\textcolor{orange}{\textbf{circles}} | \textcolor{blue}{\textbf{squares}}: the facet values from the participants facet questionnaire for \textcolor{orange}{Abi-like} and \textcolor{blue}{Tim-like} facet values respectively; 
\textcolor{orange}{\textbf{square outline}} | \textcolor{blue}{\textbf{square outline}}: \textcolor{orange}{Abi-like} | \textcolor{blue}{Tim-like} facet values participants expressed when they ran into a bug.  }
\vspace{3mm}
\label{fig:bug1&2}
\end{figure}

\textcolor{black}{Team~F localized the fault behind Bug~2 (IA \textit{wheres}) to the issue list's} nondescript titles, uninformative descriptions, and limited labeling.
Team~F realized that, with this IA, the Issue List gave little indication as to whether an issue would fit a newcomer's skill level \textit{(Label-IndexTerm, Label-Header)}.
Stage~Three proved Team~F to be right: Bug~1 \& 2 did affect several participants (Figure \ref{fig:bug1&2}):

\longquote{P1-O}{...I don't really know...I would say if I had to fix [an issue from the issue list], I'd probably just ask someone for help.}

\boldification{ In particular, as predicted by team F these bugs were more problematic for Abi-like participants}

\begin{figure}[!tp]
\graphicspath{ {figures/} }
\centering
         \includegraphics[width=.47\textwidth]{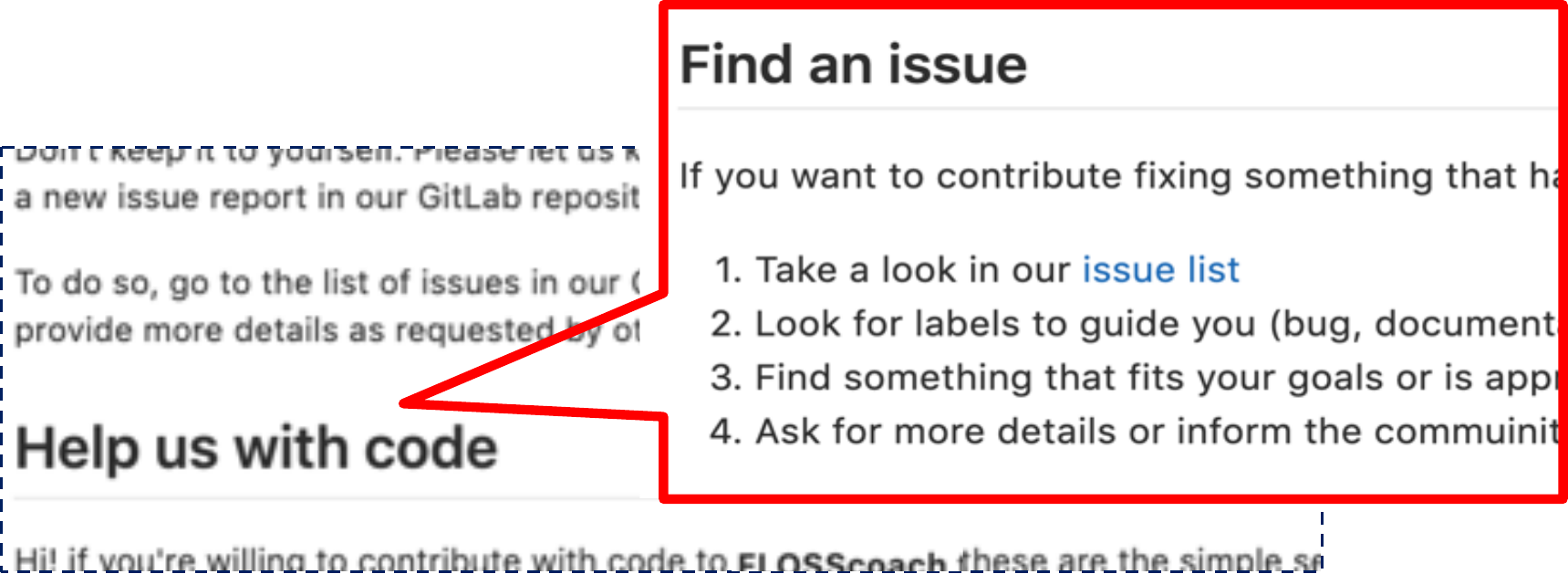}
\caption{Bug~1 before the fix, the screen appeared as shown without the call-out, giving little guidance on how to find a suitable issue. The fix added the ``Find an issue'' process description.}
\vspace{2.5mm}
\label{fig:FindIssue}
\end{figure}

\boldification{a label link was changed... a section added following a particular org scheme...  }
To fix Bug~1 \textcolor{black}{(Table~\ref{table:FlossIssuesFixes} rol 1 col. 3)}, Team~F made several changes to the IA. They created better cues for the link to the contribution guidelines by changing its label \textit{(Label-ContextualLink)} from the file name (``contributing.md'') to ``contributing guidelines'' and including additional keywords about what to expect from the link. They also modified the IA of the ``contributing.md'' to point out specific task-oriented instructions for finding an issue \textit{(OrgScheme-Task)} including a header \textit{(Label-Header)}--``Find an issue'' (Fig~\ref{fig:FindIssue}), a link to the ``issue list''\textit{(Nav-ContextualLink, Label-ContextualLink)}, and additional keywords \textit{(Label-IndexTerm)} to \textcolor{black}{add support for process-oriented} and risk-averse newcomers.

\begin{figure}[ht!]
\centering
\vspace{-4mm}
   \includegraphics[width=0.94\linewidth]{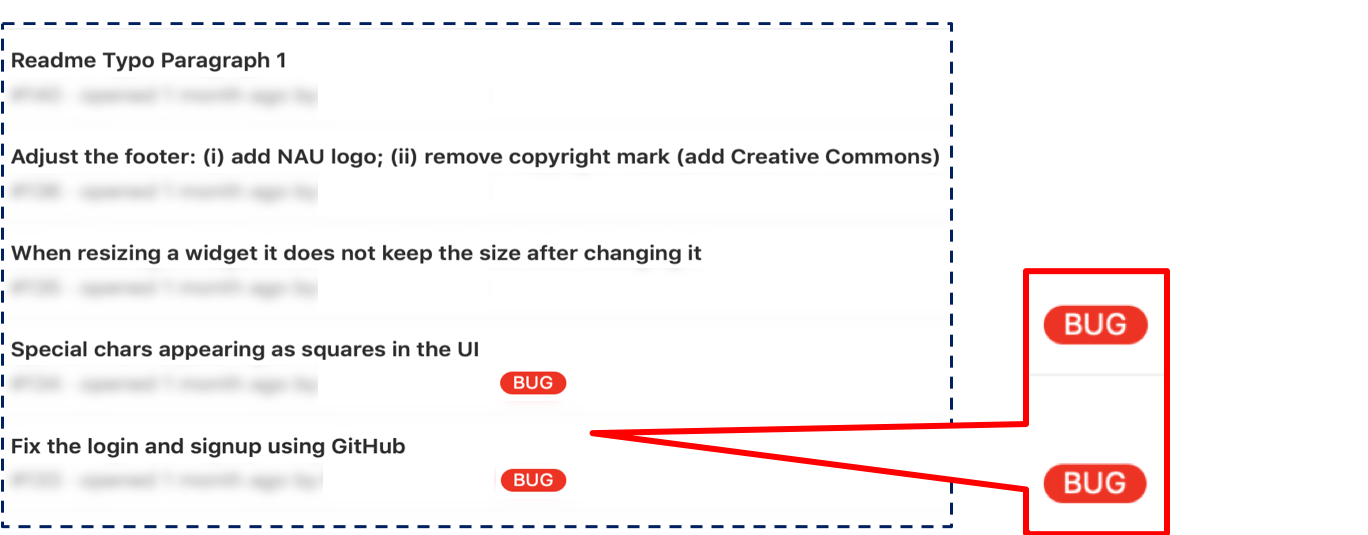}
   \includegraphics[width=0.94\linewidth]{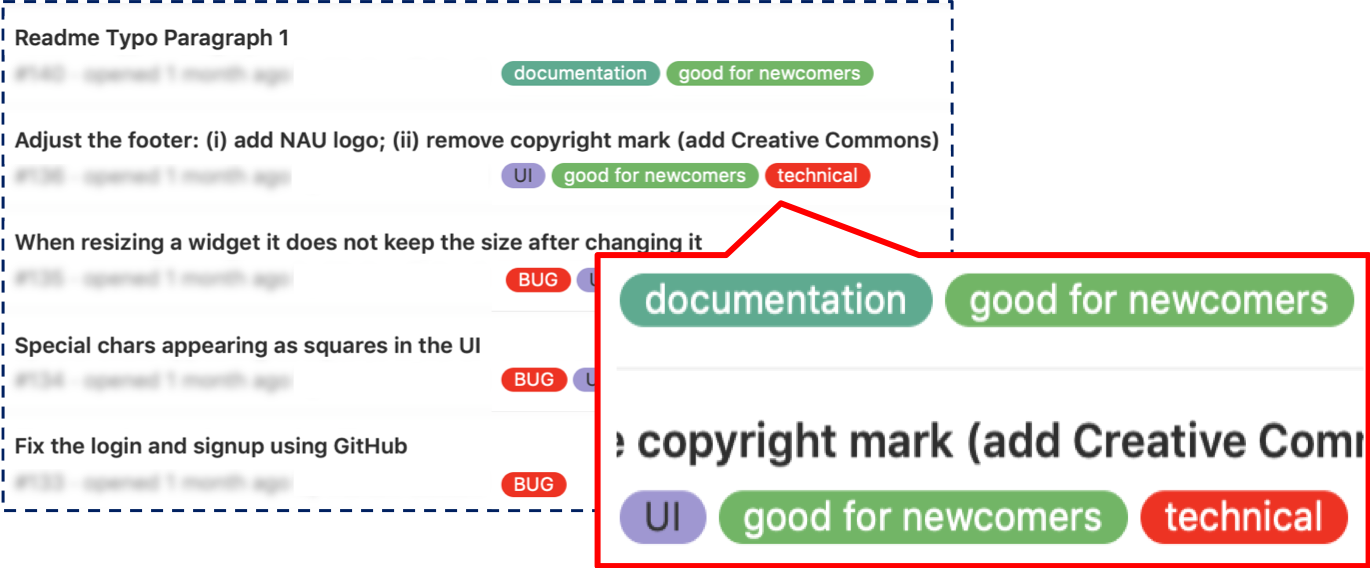}
\caption{Top: Bug~2 before the fix had only one label (``Bug''). Bottom: The fix added multiple descriptive labels.}
\label{fig:BeforeAfterLabels}
\end{figure}

Team~F fixed Bug~2 (Table~\ref{table:FlossIssuesFixes} row 2 col. 3) with improved issue headers and labels \textit{(Label-Header, Label-IndexTerm).} 
The labels signaled attributes of the open issues in the project (Figure \ref{fig:BeforeAfterLabels}).  
\textcolor{black}{Team~F also rewrote some issue descriptions to support newcomers with a comprehensive information processing style. } 

\boldification{These fixes truly had an impact on participants deciding which issue they would feel comfortable fixing.  Here is the story of P17-D...}
In Stage~Three, the OSS newcomer participants showed 
that Bug 1 \& 2 were pervasive; \emph{all} participants using the Original version faced problems related to Bug 1 and/or 2 (Figure \ref{fig:bug1&2}). 
But were these bugs \textit{inclusivity} bugs, i.e., \textit{disproportionately} affecting people with particular cognitive styles?

Figure \ref{fig:bug1&2} answers this question.
Counting up the colored outlines, which show the facets Stage~Three participants verbalized \textit{when they ran into those bugs}, shows that Bug~1 \&~2  disproportionately impacted Abi-like facet values: 74\% (14/19) of the facets participants verbalized with Bug~1 \&~2  were Abi-like facet values
(orange square outlines in Figure \ref{fig:bug1&2}, left). 

Although Bug 1 \& 2 disproportionately affected \textcolor{black}{participants with} Abi-like facet values, targeting these facets helped  \textcolor{black}{participants} across the entire cognitive style spectrum, \textit{both for Abi-like and Tim-like newcomers} (Figure~\ref{fig:bug1&2}, right). 
Further, only one participant of the DiversityEnhanced treatment (P15-D, Figure \ref{fig:bug1&2}) ran into these bugs---compared to all 9 participants in the Original treatment (Table~\ref{table:UserStudyGeneral}).

Even when participants veered off track, the  \textcolor{black}{label fixes} 
\textit{(Label-IndexTerm)} (Figure~\ref{fig:BeforeAfterLabels}) helped them find their way back. For example, P17-D initially chose an issue labeled ``good for newcomers'' and ``technical'', but soon found that they would have needed more coding experience. 
P17-D realized that issues that did not include the ``technical'' label would be a better fit.

\longquote{P17-D}{...and in fear of not making the same mistake, I'm just going to go with a [issue], which only says good for newcomers and documentation.}
%-------------------------------------
\subsection{Bug 3: ``I would expect something linear''}

\boldification{Team~F felt 4 Abi facets impacted for those wanting to make doc contributions P2-O quote.}
When evaluating the documentation contribution use-case (U2-Doc-ument), Team~F predicted that newcomers might think that they have to go through all the technical setup in order to make any contribution, even a documentation contribution (Bug~3). 
Team~F's \textit{why} analysis \textcolor{black}{(Table~\ref{table:FlossIssuesFixes}'s third row)} pointed to four of Abi's cognitive styles: task-oriented motivations, process-oriented learning, relatively low self efficacy, and risk aversion. 
Team~F localized Bug~3's fault in the IA (\textit{wheres}) to point to the absence of keywords that could guide newcomers in contributing documentation. 

\begin{figure}[!b]
\graphicspath{ {figures/} }
\centering
         \includegraphics[width=.47\textwidth]{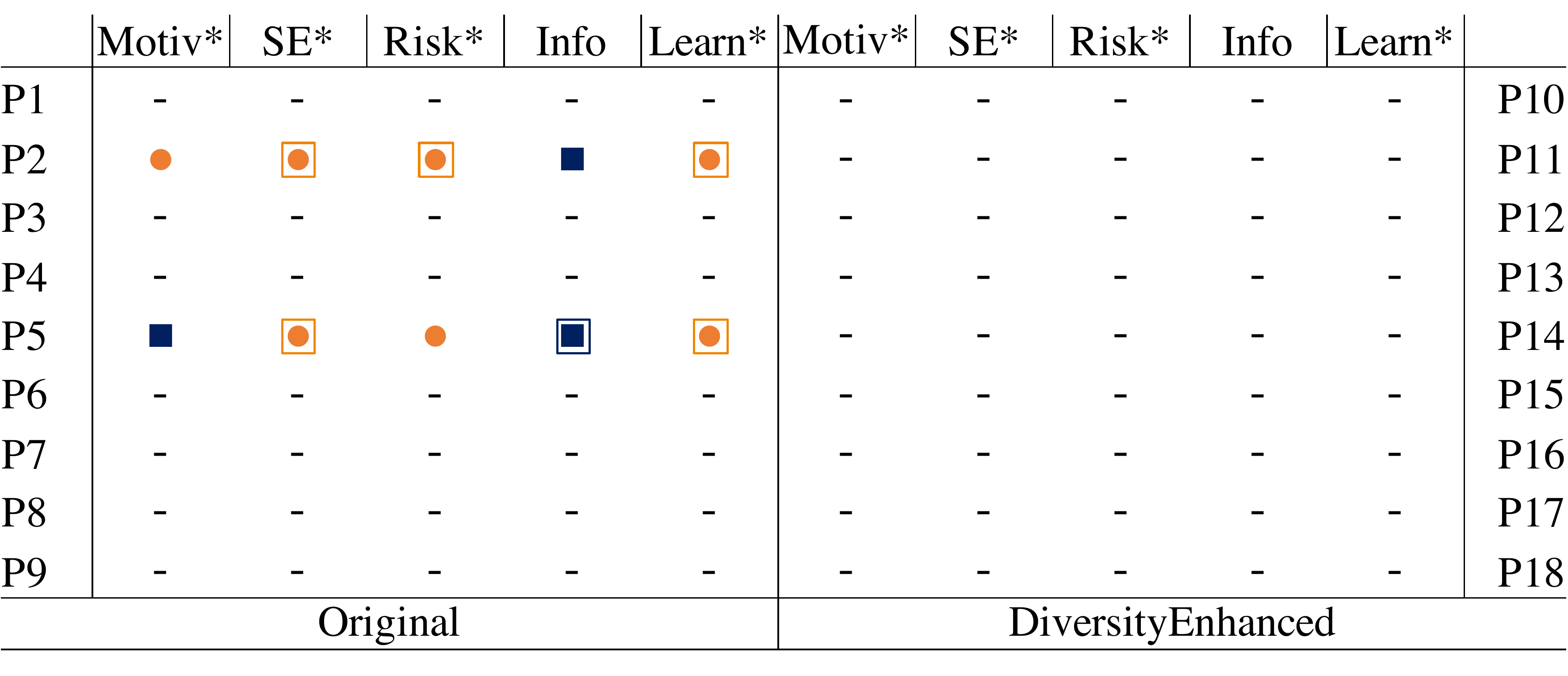}
\caption{Two Original treatment participants ran into Bug 3, but nobody using the DiversityEnhanced version did.
*, \textcolor{orange}{\textbf{circles}}, \textcolor{blue}{\textbf{squares}}: see Figure~\ref{fig:bug1&2}.}
\label{fig:bug3}
\end{figure}

In Stage~Three, Team~F's prediction was borne out: 
two lab participants did run into Bug~3 (Figure~\ref{fig:bug3}). For example: 
\longquote{P2-O (risk-averse as per facet questionnaire responses)}{Should I be doing this? Like, should I be coding just to change an N to an M? Seems a little unnecessary?...I'm stuck.}

The lack of a task-centric organization scheme for the instructions also impacted P2-O, \textcolor{black}{ a process-oriented learner according to their facet questionnaire responses:} 
\longquote{P2-O}{I would expect something linear.}

\boldification{Project~F's changed the IA this way and targeted all 4 facets (Motiv, SE, risk, and learn) when creating fixes. Here's how...}

As Table~\ref{table:FlossIssuesFixes} row 3 col. 3 summarizes, Team~F fixed the IA by mentioning ``contributions with documentation'' in the \texttt{README.md} \textit{(Label-IndexTerm)}, and by organizing \texttt{contributing.md} information  with a header \textit{(Label-Header)} that followed a task-based organization scheme \textit{(OrgScheme-Task)}, to support people with Abi-like motivations. 
Team~F also added step-by-step instructions, keywords \textit{(Label-IndexTerm)} and links to detailed information \textit{(Nav-ContextualLink)}, to support diverse learning and information processing styles.

The results of Stage 3 showed that the changes had positive effects. 
As Figure~\ref{fig:bug3} shows, although two participants ran into Bug~3 with the Original version, nobody did using the DiversityEnhanced version. 

%----------------------------------------------
\subsection{Bug~4: Where to go to file an issue}
For Bug~4 Team~F decided that, in trying to file an issue (use-case U3-FileIssue), newcomers might not know where to go, especially those who are risk-averse, those with comprehensive information processing styles or relatively low self-efficacy (\textcolor{black}{Table~\ref{table:FlossIssuesFixes} row 4 col. 1).} The elements of IA \textit{where} the team found these problems were in \textit{Nav-ContextualLink, Label-IndexTerm}, and \textit{Label-ContextualLink} elements.

However, Team~F was wrong---\textcolor{black}{in Stage~Three}, none of the Original version lab participants ran into Bug~4.
The reason was a flaw in Team~F's analysis of this use-case as it related to newcomers' prior experience.
In the Stage~Three task sequence, participants had already been to the ``issue list'' in context of an earlier use-case (U1-Find).
Thus, as P5-O put it:
\longquote{P5-O}{Since I already spent some time on that issue page [issue list]. That part [filing an issue] was not too hard.}

Still, Stage~Three had not yet occurred, and Team~F made the IA fixes in Stage~Two to fix the bug 
(Table~\ref{table:FlossIssuesFixes} row 4 col. 3).
The Stage~Three participants who then used the DiversityEnhanced version experienced no problems.
Thus, the question of whether newcomers \textit{would have} run into these problems if they had not previously learned the features remains unanswered.
However, the question of whether newcomers ran into problems in the changed version is answered: nobody ran into any problems in the DiversityEnhanced version (Table~\ref{table:UserStudyGeneral}).

%------------------------------------------
\subsection{Bug 5 \& 6: What, where, and how to set up}

\boldification{Two bugs, and here's Bug 5: Abi would not be able to find the correct instruction for her OS}
In use-case U4-Setup, Team~F's analysis revealed Bug~5 (\textcolor{black}{Table~\ref{table:FlossIssuesFixes}'s fifth row), namely that newcomers with comprehensive information processing style, low self-efficacy, or risk aversion could run into problems finding the setup instructions for their particular operating system (OS). Team~F identifed the underlying faults to be the \textit{Label-IndexTerm}, \textit{OrgScheme} and \textit{OrgStruct}, none of which were pointing out where different OSs' setup instructions might be. 
} 

\boldification{And here's Bug~6: she would not know where to put the commands and would want more explanation about the commands}
Even if a newcomer overcame Bug~5 and found the \textcolor{black}{right instructions, Team~F 
realized that an OSS newcomer might not necessarily ``just know'' what each command in the instructions actually did or exactly where to run them (Bug~6: Table~\ref{table:FlossIssuesFixes}'s sixth row).  
As the table shows, Team~F's \textit{why} analysis suggested that this inclusivity bug could particularly affect a newcomer with \textit{any} of Abi's cognitive style values, due in part to the absence of hints with clarifying keywords (e.g., ``command line terminal...'') \textit{(Label-IndexTerm)}.
} 

\boldification{And yep -- these problems were indeed problems.}

\begin{figure}[!bp]
\graphicspath{ {figures/} }
\centering
         \includegraphics[width=.47\textwidth]{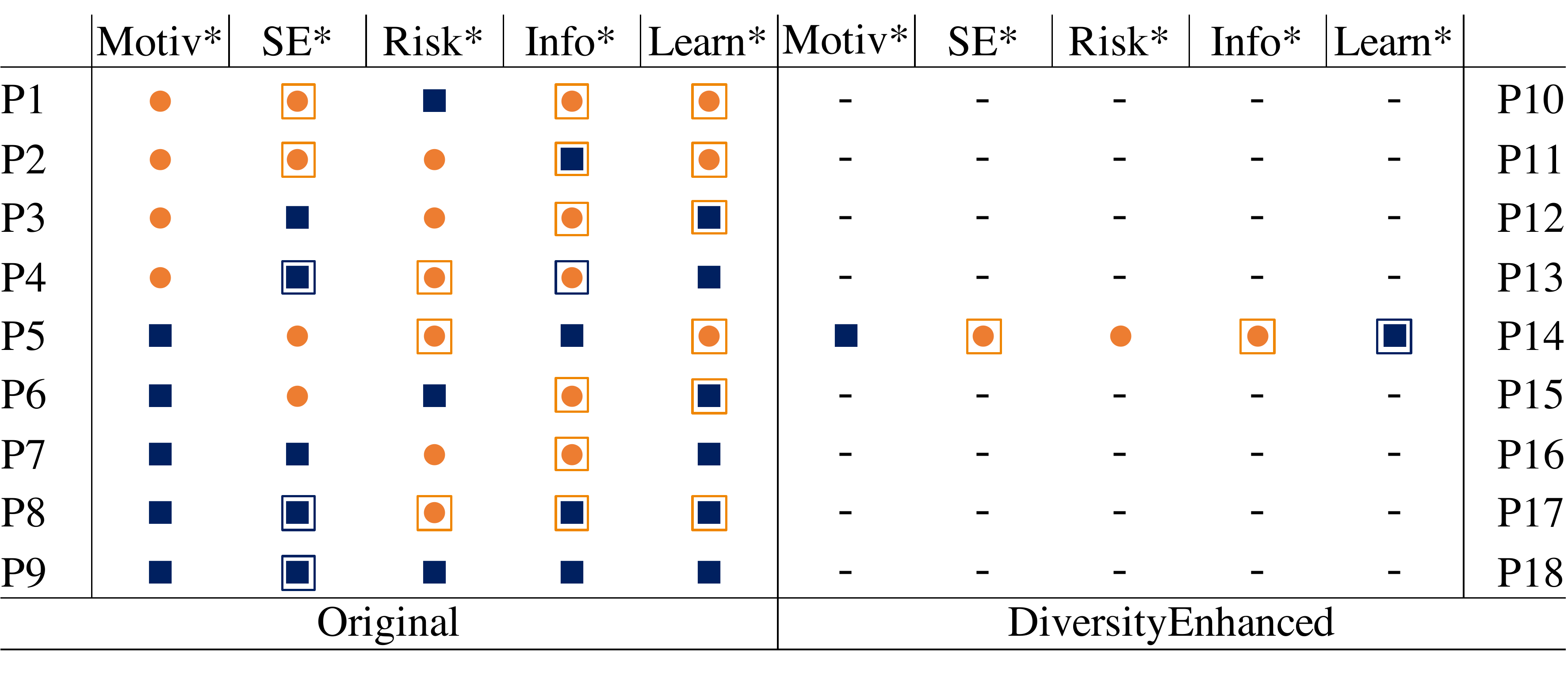}
\caption{All Original participants but only 1 DiversityEnhanced participant ran into Bug 5 \& 6. *, \textcolor{orange}{\textbf{circles}}, \textcolor{blue}{\textbf{squares}}: see Figure~\ref{fig:bug1&2}.}

\label{fig:bug5&6}
\end{figure}

Stage~Three's results confirmed Team~F's predictions: all Original participants ran into one or both of these bugs (Figure \ref{fig:bug5&6}).
Also as per Team~F's prediction, when  participants  ran into the bugs, they verbalized mostly Abi-like facet values: for Bug~5 \& 6, 81\% (17/21) were Abi-like facet values (orange square outlines left half Figure~\ref{fig:bug5&6}). For example:
\longquote{P1-O (low-self-efficacy)}{I feel like they [the OSS developers] put up barriers because they would want people that really knew what they were doing...} % about Bug 5
\longquote{P1-O (continues)}{I'd probably just, like, not work on it.}

\begin{figure}[!b]
\centering
   \includegraphics[width=0.94\linewidth]{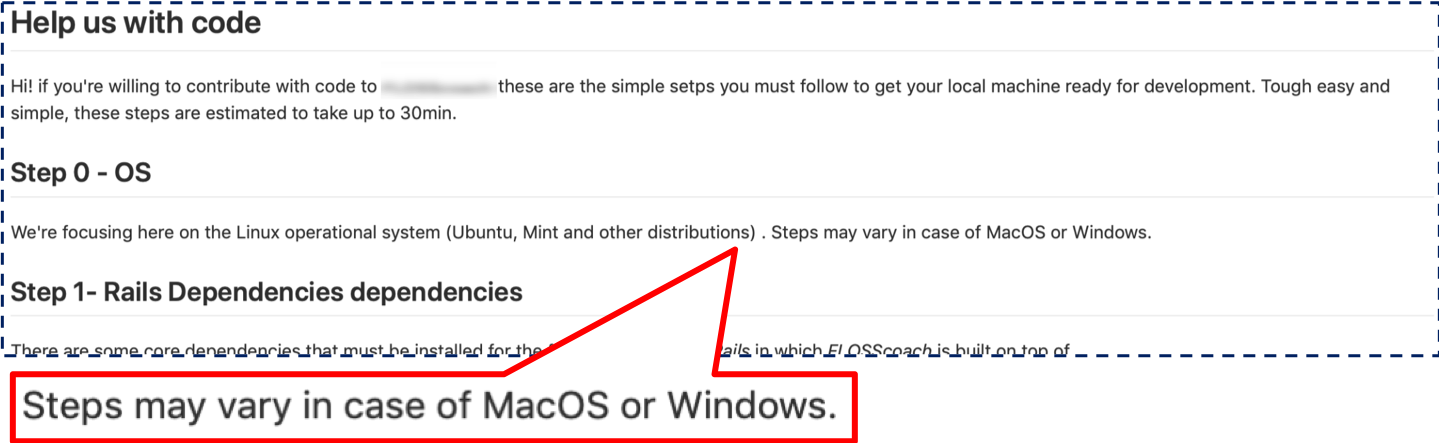}
   \includegraphics[width=0.94\linewidth]{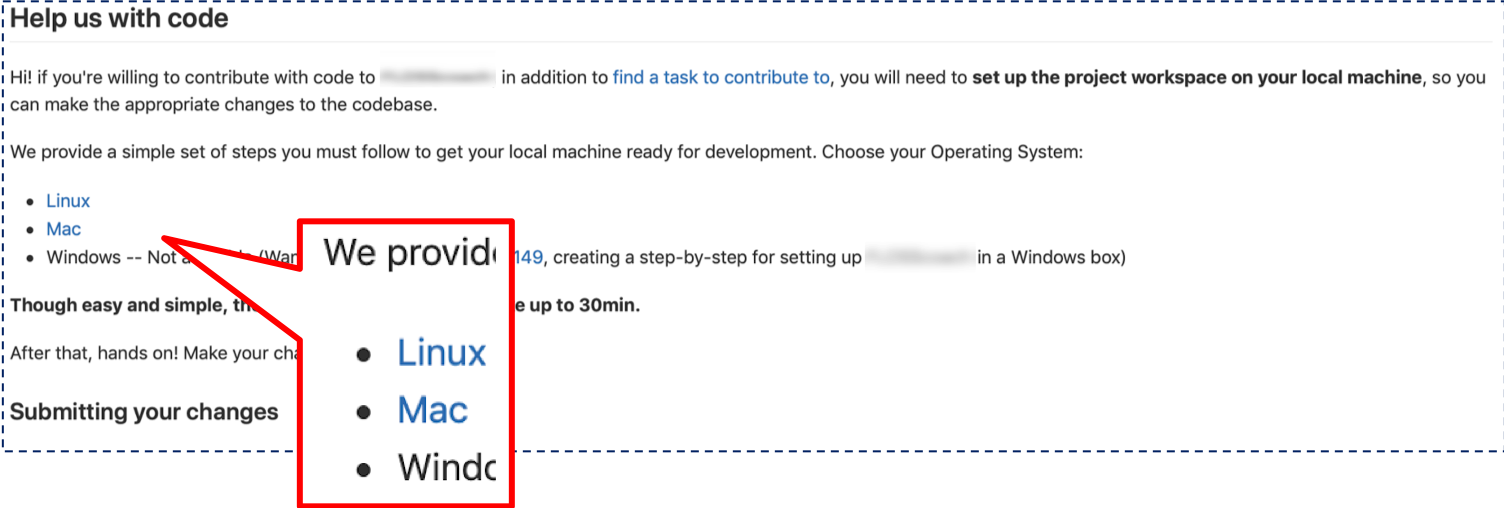}
\caption{Top: Bug~5 before the fix: no scheme or cues/keywords to enable finding instructions for different OS's. Bottom: Bug~5's fix added topic-based scheme and linked to instructions for each OS.}
\label{fig:BeforeAfterEnv}
\end{figure}

The lab participants also pointed out mismatches for cognitive styles like process-oriented learning, comprehensive information processing, and risk-aversion to using commands they did not completely understand:
\longquote{P1-O}{These instructions aren't working super good for me ... if there was explanations a little more.} 
\longquote{P3-O}{I don't completely understand ... where to move it [a command] or where to put it.} 
 
\boldification{Different links were added for each OS system setups and a hint in the readme}

To address Bug~5, Team~F restructured the ``Help us with code'' section by adding a layer of hierarchy to structurally identify general information about code contributions \textit{(OrgStruct-Hierarchy)}. They also reorganized the section topically by OS type \textit{(OrgScheme-Topic)} (Figure~\ref{fig:BeforeAfterEnv}). Moreover, they added keywords \textit{(Label-IndexTerm)} in the \texttt{README.md} similar to Bug~3's fix, to more clearly guide newcomers to the right setup instructions for their OS.
To fix Bug~6, Team~F added explanations to each step in the instructions, in which they made explicit the reason for each step and the need to use a command line terminal for the commands \textit{(Label-IndexTerm)}.

Team~F's IA fixes paid off: both Abi-like and Tim-like participants improved and the number of participants who ran into problems decreased from 9 to 1, an 89\% improvement (Figure~\ref{fig:bug5&6}).
Further, although \textit{none} of the Original participants completed the task successfully, \textit{all} participants using the DiversityEnhanced version were able to complete the task---even P14-D, who at first ran into a problem, but overcame it and eventually succeeded. 
\section{Discussion}
\label{sec:discussion}
\subsection{The IA Fixes: Equity and Inclusion}

\boldification{Question: So, these fixes helped a lot, but -- what about equity? And what about inclusion?}

As the results sections have shown, the IA fixes that differentiated the DiversityEnhanced version from the Original version led to a 90\% reduction in the bugs that Team~F had found to be inclusivity bugs (Section~\ref{sec:results_study2and3}'s Table~\ref{table:UserStudyGeneral}). 
However, this leaves unanswered whether these fixes actually contributed to the goals of making the project's infrastructure (1)~more \textit{equitable} and (2)~more \textit{inclusive}. 
For example, equitability could be achieved by helping one group at the expense of another, but that would not achieve inclusivity.
Team~F's goal was to do both.

\boldification{1. Q: equity? A: Yes. how we know: num orange vs \review{blue} outlines Orig:inequitable.  DivEnhanced: pretty equitable}

First we consider equity.  
A dictionary definition of equity is ``the quality of being fair and impartial''~\cite{dictionary}.
We measured equity analyzing the lab participants' data, because the participants covered an almost equal number of Abi and Tim facets (recall Figure~\ref{fig:facetpergroup}: 22 Abi facet values and 23 Tim facet values in each treatment). Thus, if the lab participants' number of ``Abi facets'' affected by a bug was greater than the number of ``Tim facets'', or vice-versa, we  conclude that the bug was inequitable in the ways it affected the participants. 
\begin{table}[!b]
%\vspace{-0.58cm}
  \caption[caption]{Inclusivity summary: Team~F's IA fixes' effects on the \textcolor{orange}{Abi}-like facet values (top) and
  the \textcolor{blue}{Tim}-like facet values (bottom) were all positive, showing that the IA fixes increased the inclusivity of the prototype across all cognitive styles.
\\+:More successes in Version DE; -:fewer
(zero occurrences).
Grayed out: nobody with these facets ran into this bug.
}
\vspace{-2mm}

\begin{tabular}{L{2cm}C{1cm}C{1cm}C{1cm}C{1cm}C{1cm}}
  \hline
   {    \textbf{Bug ID} }
    & {   \textbf{Motiv}}
    & {   \textbf{SE} }
    & {   \textbf{Risk} }
    & {   \textbf{Info} }
    & {   \textbf{Learn} }
   \\\hline

%---------Abi facets start here------------
 
     Bug 1 \& 2 &    \textcolor{orange}{+} & \textcolor{orange}{+} & \textcolor{orange}{+} & \textcolor{orange}{+} & \textcolor{orange}{+} \\ 
     %\hline  

    Bug 3&    \textcolor{orange}{+} & \textcolor{orange}{+} & \textcolor{orange}{+} & 
    \cellcolor{gray!15} \textcolor{orange}{ } & \textcolor{orange}{+} \\ 
    %\hline  
  
  %   Issue 4&    9/9 &    9/9 \\ %\hline  
  
     Bug 4&    \cellcolor{gray!15} \textcolor{orange}{ } & \cellcolor{gray!15} \textcolor{orange}{ } & 
     \cellcolor{gray!15} \textcolor{orange}{ } & 
     \cellcolor{gray!15} \textcolor{orange}{ } &
     \cellcolor{gray!15} \textcolor{orange}{ } \\ 
     %\hline
  
  %   Issue 6&    9/9 &    7/9 \\ %\hline  

     Bug 5 \& 6 &    \textcolor{orange}{+} & \textcolor{orange}{+} & \textcolor{orange}{+} & \textcolor{orange}{+} & \textcolor{orange}{+} \\ 
     %\hline  
 %   \textbf{Total issues experienced} &    \textbf{20} &    \textbf{2}\\ 
 %   \textbf{Mean}}&    \textbf{5}} &    \textbf{1.75}}\\ 
 \hline
 
 %---------Tim facets start here------------
      Bug 1 \& 2 &    \textcolor{blue}{+} & \textcolor{blue}{+} & \textcolor{blue}{+} & \textcolor{blue}{+} & \textcolor{blue}{+} \\ 
     %\hline  

    Bug 3&    \textcolor{blue}{+} & 
    \cellcolor{gray!15} \textcolor{blue}{ } & 
    \cellcolor{gray!15} \textcolor{blue}{ } & \textcolor{blue}{+} & 
    \cellcolor{gray!15} \textcolor{blue}{ } \\ 
    %\hline  
  
  %   Issue 4&    9/9 &    9/9 \\ %\hline  
  
     Bug 4&    \cellcolor{gray!15} \textcolor{blue}{ } & \cellcolor{gray!15} \textcolor{blue}{ } & 
     \cellcolor{gray!15} \textcolor{blue}{ } & 
     \cellcolor{gray!15} \textcolor{blue}{ } & 
     \cellcolor{gray!15} \textcolor{blue}{ } \\ 
     %\hline
  
  %   Issue 6&    9/9 &    7/9 \\ %\hline  

     Bug 5 \& 6 &    \textcolor{blue}{+} & \textcolor{blue}{+} & \textcolor{blue}{+} & \textcolor{blue}{+} & \textcolor{blue}{+} \\ 
     \hline
  \end{tabular}

  \label{table:Effects_on_personas}

\vspace{-3.5mm}
\end{table}

%--------------------OLD VERSION (lumps all facets into 1 persona)
%\begin{table}[!hbt]
%\begin{tabular}{L{4.25cm}C{2cm}C{3.25cm}}
%  \hline
%   {    \textbf{Issue ID} }
%    & {   \textbf{Abi}}
%    & {   \textbf{Tim} }
%    \\\hline
%
%     Issue 1 \& 2 &    \textcolor{orange}{+} &    \textcolor{blue}{+} \\ %\hline  
%
%    Issue 3&    \textcolor{orange}{+} &    \textcolor{blue}{*} \\ %\hline  
%  
%  %   Issue 4&    9/9 &    9/9 \\ %\hline  
%  
%     Issue 5&    \textcolor{orange}{*} &    \textcolor{blue}{*} \\ %\hline
% % 
%  %   Issue 6&    9/9 &    7/9 \\ %\hline  
%
%     Issue 7 \& 8 &    \textcolor{orange}{+} &    \textcolor{blue}{+} \\ %\hline  
 %   \textbf{Total issues experienced} &    \textbf{20} &    \textbf{2}\\ 
 %   \textbf{Mean}}&    \textbf{5}} &    \textbf{1.75}}\\ 
 %\hline
 % \end{tabular}
 % \caption{\textcolor{red}{MMB: don't use this table, it's very misleading -- too much got averaged, and much information got lost.}Each row shows how Team~F's fixes affected participants' \textcolor{orange}{\textbf{Abi}} and \textcolor{blue}{\textbf{Tim}} facet values. Issues in the same use-case are grouped together.
%\\ \textbf{+}: Fewer Abi-like or Tim-like participants ran into issues in %Version DE than in Version O.
%\textbf{*}: No effect because nobody ran into Issues.
%  } 
%  \label{table:Effects_on_personas}
%
%\vspace{-4mm}
%\end{table}

By this measure, Bugs 1 \& 2 in the Original version were inequitable: together they affected 14 of participants' Abi facets (orange outlines for Figure~\ref{fig:bug1&2}'s Original version), compared to only 5 Tim facets (blue outlines).
Applying the same measure to the DiversityEnhanced version shows that, although the DiversityEnhanced version was still slightly inequitable---two of participants' Abi facet inequities (2 orange outlines), and zero Tim facet inequities---it was less inequitable than the Original version.
Applying the same measures to Bug 3 (Figure~\ref{fig:bug3} - Original: 5~Abi/1~Tim; DiversityEnhanced: 0~Abi/0~Tim) and to Bugs 5 \& 6 (Figure~\ref{fig:bug5&6} - Original: 17~Abi/4~Tim; DiversityEnhanced 2~Abi/1~Tim) also show that the IA fixes likewise reduced the inequities. Thus, we can conclude that the IA fixes did make Project~F's infrastructure more equitable.  

\boldification{2. Q: inclusion? (Did we trade away Tim in order to help Abi?) A: Yes, more inclusion. how we know: Num orange outlines Decreased AND Num \review{blue} outlines Decreased, so more inclusive across both. As Table~\ref{table:Effects_on_personas} summarizes, every facet for everyone got better.}
Inclusion can be computed using a different measure on the same data. 
According to the dictionary, inclusion is ``the action or state of including or of being included within a group or structure''~\cite{dictionary}.
Applying this definition to being included by a bug fix, we will conclude that the bug fix was inclusive if the number of lab participants' facets affected by a bug decreased from the Original version to the DiversityEnhanced version for participants' Abi facets \textit{and} for participants' Tim facets.

Applying this measure to Bugs 1 \& 2 (Figure~\ref{fig:bug1&2}) reveals that, after the fix, participants' Abi facets affected decreased by 12 (from 14 facets affected to 2). 
Likewise, participants' Tim facets affected decreased by 5 (from 5 facets affected to 0).
Since the number of participants' facets affected decreased for participants' Abi facets \textit{and} for participants' Tim facets, we conclude that the fixes improved inclusivity.
Applying the same measures to Bug 3 (Figure~\ref{fig:bug3} - Abi: 5~Original/0~DiversityEnhanced, Tim: 1~Original/0~DivEnhanced) and Bugs 5 \& 6 (Figure~\ref{fig:bug5&6} - Abi:  17~Orig/2~DivEnhanced, Tim: 4~Orig/1~DivEnhanced) shows that they also improved inclusivity.
As  Table~\ref{table:Effects_on_personas} shows, for every bug and every facet value, participants' Abi-facets and Tim-facets all ran into fewer barriers in the DiversityEnhanced version.   

%----------------------------------------------
\subsection{What about gender?}
\begin{figure}[!b]
\graphicspath{ {figures/} }
\centering
      \includegraphics[width=.45\textwidth]{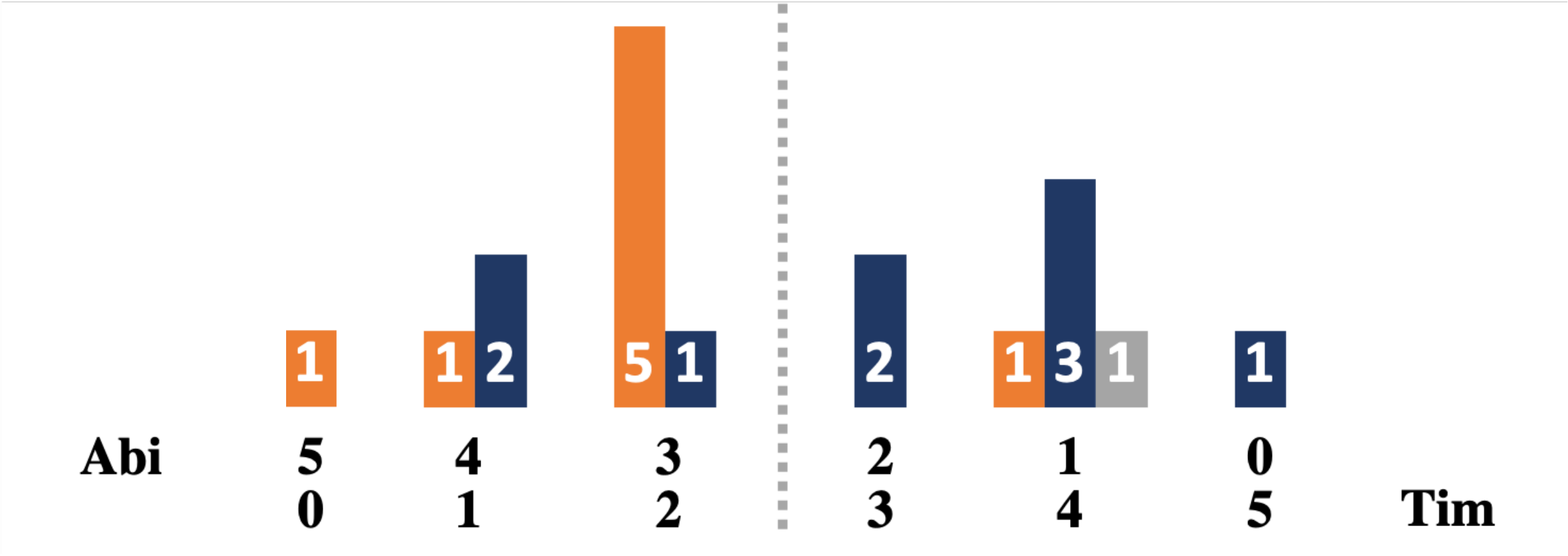}
\caption{\# of women (orange), men (black), and decline-to-specify (gray) with each combination of facets (from facet questionnaire), using the same x-axis scheme (from 5 Abi facets to 5 Tim facets) as Figure \ref{fig:PersonaDistribution}. Note that the right half of the graph contains only 1 of the 8 women participants.
}
\label{fig:genderdistribution}
 \end{figure}

In some prior literature (e.g., \cite{vorvoreanu-chi19}), analyses of these cognitive styles have revealed gender differences.
That was also the case for our Stage~Three participants' cognitive styles. 
The participants displayed a range of facet values, but as in other studies, women's facet values tended more ``Abi-wards'' than the other participants' (Figure \ref{fig:genderdistribution}).
These results agree with previous literature that explain how these facets tend to cluster by gender~\cite{burnett2016gendermag}. 
\textcolor{black}{These results also, when taken together with Figure~\ref{fig:bug1&2}, Figure~\ref{fig:bug3}, and Figure~\ref{fig:bug5&6}, show that most of the facets affected by the bugs were those of the women participants.}

\textcolor{black}{However, the SUS usability ratings did not differ much by gender.
First, as Table~\ref{table:SUS_Results} shows, the SUS scores of participants who used the Original project were equally low across gender,
which may suggest that the Original had a long way to go from everyone's perspective. 
Second, the SUS scores for participants who used the DiversityEnhanced project were much higher across gender, adding to the body of evidence (e.g., \cite{ljungblad2007transfer, vorvoreanu-chi19}) that designing for often-overlooked populations (here, Abi) can benefit everyone. 
}

\begin{table} [!t]

\centering
%\vspace{-3 mm}
  \caption{Participants' SUS rating scores. (Maximum possible for the subset we used: 32.)}
  \vspace{-2 mm}
\begin{tabular}{L{3.4cm}C{2.6cm}C{3.1cm}}
  \hline
    & {  \textbf{Original}} 
    & {  \textbf{DiversityEnhanced}}
    \\\hline
    
      Men's Average
    &   12 (6 Men)
    &   19 (3 Men)
    \\
    
    {Women's Average}
    &   12 (3 Women)
    &   22 (5 Women)
    \\
    
    {Gender-not-stated}
    &   N/A
    &   32
    \\
    
    { Overall Average}
    &   12
    &   22
    \\\hline
    %\\
    
%    {  SD}
%    &{  6.05}
%    &{  6.78}
%    \\
    
    %{  Median}
    %& {  11}
    %& {  21}
    %\\\hline
  \end{tabular}

\label{table:SUS_Results}
\vspace{-3mm}
\end{table}
\subsection{The Facet Questionnaire's Validity}
\boldification{The facet questionnaire is used by researchers and practitioners...But does it align with participants facets' behavior?}

As a few other researchers have also done~\cite{gralha2019, hilderbrand2020gendermag-bp, vorvoreanu-chi19}, we used the cognitive facet questionnaire (Section~\ref{subsec:stageThree}) 
to collect the participants' facet values.
However, we also collected facet values from a second source: participants' verbalization during their tasks.  
These two sources enabled us to consider the consistency of the questionnaire's responses with the facets that actually arose \textcolor{black}{among the participants}. 

The data comparing \textcolor{black}{participants'} facet questionnaire responses with \textcolor{black}{their} actual \textit{in-situ} facet occurrences were detailed earlier in 
Figure~\ref{fig:bug1&2}, Figure~\ref{fig:bug3}, and Figure~\ref{fig:bug5&6}. 
Outline colors depict the \textit{in-situ} facet occurrences that arose; the shape's fill color depicts the participant's questionnaire response for that facet.
(No outline color simply means no evidence arose \textit{in-situ} about that facet.)
Thus, when an outline color matches the shape's fill color (questionnaire response), then the questionnaire captured that participant's facet value correctly for the situation.

\boldification{Turns out the facet questionnaire does a good job. In 78\% of the cases the verbalized facets aligned with the facet survey responses} 

Overall, 78\% of participants' \textit{in-situ} facet verbalizations aligned with their facet questionnaire responses which suggests that the facet questionnaire was a reasonable measure of participants' facet values. 

\section{Threats to validity}
\label{sec:threat}
As with any empirical research, our investigation has threats to validity. In this section, we explain threats related to our investigation and ways we guarded against them. 

{\color{black}During Stage~One, Team~F reported the issues found in their project from the perspective of one type of newcomer based on GenderMag's Abi persona. 
Past research has suggested using the Abi persona first~\cite{hilderbrand2020gendermag-bp}, since Abi's facet values tend to be more undersupported in software than those of the other personas (e.g., \cite{burnett-fieldstudy-2016}). 
However, fixing problems from only this persona's perspectives could leave non-Abi-like newcomers less supported than before. 
We mitigated this risk by empirically evaluating the fixes with both Abi-like and Tim-like newcomers.
That said, some cognitive facets are not considered at all by GenderMag personas, such as memory or attention span, which
could be particularly pertinent to people with even mild cognitive disorders. Our investigation did not account for those types of cognitive facets.}

As with any investigation with a lab study component, we needed to choose a setting, and our setting (OSS Project~F) may not generalize to other OSS projects.
The relatively small number of participants (18 in total), which was necessary for tractability of qualitative analysis, also threatens generalizability.
In addition, our Stage~Three investigation could have uncontrolled differences between the two participant groups. {\color{black}To partially mitigate this threat, we used participants' facet questionnaire responses to assign them to treatments with identical facet distributions (recall Fig~\ref{fig:facetpergroup}).}

In Stage~Three, the identical sequence of the tasks (use-cases), which reflects a workflow common for OSS contributions~\cite{Steinmacher.ea_2016}, may have created learning effects that could have influenced the results.
Finally, our comparison of facet questionnaire results against verbalizations had only partial data available, since we coded facets from only participants' verbalizations when they encountered a bug, and P5-O's audio for Bug~1 \& 2 were corrupted, so we only had observation notes for that participant. 

Threats like these can be addressed only by additional studies across a spectrum of empirical methods that isolate particular variables and establish the generality of findings over different types of OSS projects, populations, and other information rich-environments. 
\section{Conclusion}
\label{sec:conclusion}
\boldification{This paper has investigated debugging inclusivity bugs and the role information architecture can play.} This paper has presented Why/Where/Fix, a systematic inclusivity debugging process. Why/Where/Fix harnesses information architecture, so we also investigated how IA can create inclusivity bugs.
Our setting was an OSS project's technology infrastructure. The ``whether'' aspects of our RQ1 results revealed that IA can indeed cause inclusivity bugs.  In our investigation, the OSS newcomer participants ran into IA-related inclusivity bugs 20 times~(Table~\ref{table:UserStudyGeneral}).
Our RQ2 ``whether'' results also revealed that IA can be part of the solution. In our investigation, Team-F's IA fixes reduced the number of inclusivity bugs the participants experienced by 90\%~(Table~\ref{table:UserStudyGeneral}).

\boldification{The ``how'' results were... }
Team~F's \textit{hows} of the above results lay in the fault localization capabilities IA brought to  Why-Where-Fix:
\begin{itemize}[noitemsep,topsep=0pt,labelindent=0.2em,labelsep=0.2cm,leftmargin=*]
    \item \textit{IA and where's}: In Stage~One, Team~F localized the IA where's  
    behind the inclusivity bugs (Section~\ref{sec:results_study2and3} and Table~\ref{table:FlossIssuesFixes}), all but one which the OSS newcomers verified.  
    \item \textit{IA and fixes}: In Stage~Two, Team~F fixed the faults, by changing the IA in the ways detailed in Section~\ref{sec:results_study2and3} and summarized in Table~\ref{table:FlossIssuesFixes}. The participants in Stage~Three showed that Team~F's IA fixes helped \textit{across the cognitive diversity range} of the newcomers in our investigation (Tables \ref{table:UserStudyGeneral} and \ref{table:Effects_on_personas}).
\end{itemize}

Key to these results is that these inclusivity fixes lay not in supporting one population at the expense of another, and not in ``compromising'' to give each population a little less than they need.
Rather, as Table~\ref{table:Effects_on_personas} illustrated, the fixes produced positive effects across  diverse cognitive styles.
These results provide encouraging evidence that the Why-Where-Fix process may provide an effective way to increase the equity and inclusion of information-rich environments like OSS projects.
\begin{acks}
We thank all the study participants for their time and insight. This work is partially supported by the National Science Foundation grants 1901031, 2042324, and 2008089; DARPA grant N66001-17-2-4030; USDA-NIFA/NSF grant 2021-67021-35344; and CNPq grant \#313067/2020-1.
\end{acks}

\bibliographystyle{ACM-Reference-Format}
\bibliography{biblio}
\end{document}